\documentclass[prc,superscriptaddress,nofootinbib,noeprint]{revtex4-2}

\usepackage[english]{babel}
\usepackage[hidelinks]{hyperref}
\usepackage{dcolumn}
\makeatletter\AtBeginDocument{\let\@elt\relax}\makeatother

\usepackage{booktabs}
\usepackage{longtable}
\usepackage{graphicx}
\usepackage{amsmath}
\usepackage{float}
\usepackage[english,capitalise]{cleveref}
\usepackage{mathtools}
\usepackage{bbm}

\usepackage[utf8]{inputenc}

\usepackage{siunitx}

\usepackage{physics}
\usepackage{csquotes}

\usepackage{multirow}

\makeindex

\newcommand{\K}[1]{\ensuremath{\left(#1\right)}}
\newcommand{\Ke}[1]{\ensuremath{\left[#1\right]}}

\let\scv\v
\renewcommand{\v}[1]{\ensuremath{\boldsymbol{#1}}}

\newcommand{\mprescript}[3]{{{\vphantom{#3}}}^{#1}_{#2}\! #3 }
\newcommand{\ibraket}[4]{ \mprescript{}{#1}{ \braket{#2}{#3}}_{#4}^{}  } 
\newcommand{\ibra}[2]{ \mprescript{}{#1}{ \bra{#2} } }
\newcommand{\iket}[2]{ \ket{#1}_{#2}^{} } 
\newcommand{\iketbra}[3]{
  \ifthenelse{\equal{#3}{}}{
    \iket{#1}{#2} \ibra{#2}{#1}
  }
  {
    \iket{#1}{#2} \ibra{#2}{#3}
  }
}
\newcommand{\imel}[5]{ \vphantom{#2}_{#1}\! \mel{#2}{#3}{#4}_{#5} }

\newcommand{\p}[1][]{p^{#1}}
\newcommand{\q}[1][]{q^{#1}}
\newcommand{\pp}[1][]{p^{\prime #1}}
\newcommand{\qp}[1][]{q^{\prime #1}}

\newcommand{\rint}[1]{\int \dd{#1[]} #1[2]}

\newcommand{\reg}[2]{g_{l_#1}{\K{#2}}}

\newcommand{\de}[1]{\delta{\K{#1}}}

\newcommand{\ci}{\mathrm{i}}
\newcommand{\kcnp}{\kappa_{cnp}}

\newcommand{\kcnph}{\v{\hat{\kappa}}_{cnp}}
\newcommand{\kcnq}{\kappa_{cnq}}

\newcommand{\kcnqh}{\v{\hat{\kappa}}_{cnq}}

\newcommand{\kcnpph}{\v{\hat{\kappa}}_{cnp}^{\prime}}

\newcommand{\kcnqph}{\v{\hat{\kappa}}_{cnq}^{\prime}}

\newcommand{\oh}{\frac{1}{2}}
\newcommand{\id}{\mathbbm{1}}

\newcommand{\kd}[2]{\delta_{#1, #2}}

\newcommand{\hesix}{\textsuperscript{6}He}

\newcommand{\pq}{p,q}

\newcommand{\np}{n^\prime}

\newcommand{\kev}{\kilo \electronvolt}
\newcommand{\mev}{\mega \electronvolt}
\newcommand{\fm}{\femto \meter}

\newcommand{\btz}{B_3^{(0)}}

\newcommand{\jJcoupling}{\(\v{j}\v{J}\)-coupling}

\newcommand{\intdd}[1]{\int \dd{#1[]} }
\newcommand{\mlo}{M_{\mathrm{low}}}
\newcommand{\mlow}{\mlo}
\newcommand{\mhi}{M_{\mathrm{high}}}
\newcommand{\mhigh}{\mhi}
\newcommand{\swave}{\(s\)-wave}
\newcommand{\pwave}{\(p\)-wave}
\newcommand{\dwave}{\(d\)-wave}

\newcommand{\pnn}{p_{nn}}
\newcommand{\thpq}{\theta_{\v{p},\v{q}}}

\newcommand{\cf}[3]{\chi_{#1}^{#2}{\K{#3}}}

\newcommand{\tbi}{\mathrm{3B}}

\newcommand{\orcid}[1]{\href{https://orcid.org/#1}{\includegraphics[scale=0.055]{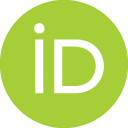}}}

\newcommand{\lbh}{\(\Lambda=\SI{1500}{\mev}\)}
\newcommand{\lbl}{\(\Lambda=\SI{1000}{\mev}\)}

\newcommand{\lgmname}{local Gaussian model}
\newcommand{\lgmnameC}{Local Gaussian model}
\newcommand{\lgmacronym}{LGM}

\begin{document}

\title{\texorpdfstring{Neutron-neutron scattering length from the $\boldsymbol{^6}$He$\boldsymbol{(p,p\alpha)nn}$ reaction}{
Neutron-neutron scattering length from the He-6(p,p alpha)nn reaction
}}

\author{Matthias G\"{o}bel \orcid{0000-0002-7232-0033}}
\email[E-mail: ]{goebel@theorie.ikp.physik.tu-darmstadt.de}
\affiliation{Technische Universit\"{a}t Darmstadt, Department of Physics, Institut f\"{u}r Kernphysik, 64289 Darmstadt, Germany}

\author{Thomas Aumann \orcid{0000-0003-1660-9294}}
\affiliation{Technische Universit\"{a}t Darmstadt, Department of Physics, Institut f\"{u}r Kernphysik, 64289 Darmstadt, Germany}
\affiliation{GSI Helmholtzzentrum f\"{u}r Schwerionenforschung GmbH, Planckstra\ss{}e 1, 64291 Darmstadt, Germany}
\affiliation{Helmholtz Research Academy for FAIR, 64291 Darmstadt, Germany}

\author{Carlos A. Bertulani \orcid{0000-0002-4065-6237}}
\affiliation{Department of Physics and Astronomy, Texas A\&M University-Commerce, Commerce, TX 75429-3011, USA}

\author{\\Tobias Frederico \orcid{0000-0002-5497-5490}}
\affiliation{Instituto Tecnol\'{o}gico de Aeron\'{a}utica, DCTA, 12.228-900 S\~{a}o Jos\'{e} dos Campos, SP, Brazil}

\author{Hans-Werner Hammer \orcid{0000-0002-2318-0644}}
\affiliation{Technische Universit\"{a}t Darmstadt, Department of Physics, Institut f\"{u}r Kernphysik, 64289 Darmstadt, Germany}
\affiliation{ExtreMe Matter Institute EMMI, GSI Helmholtzzentrum f\"{u}r Schwerionenforschung GmbH, 64291 Darmstadt, Germany}
\affiliation{Helmholtz Research Academy for FAIR, 64291 Darmstadt, Germany}

\author{Daniel R. Phillips \orcid{0000-0003-1596-9087}}
\affiliation{Institute of Nuclear and Particle Physics and Department of Physics and Astronomy, Ohio University, Athens, OH 45701, USA}
\affiliation{Technische Universit\"{a}t Darmstadt, Department of Physics, Institut f\"{u}r Kernphysik, 64289 Darmstadt, Germany}
\affiliation{ExtreMe Matter Institute EMMI, GSI Helmholtzzentrum f\"{u}r Schwerionenforschung GmbH, 64291 Darmstadt, Germany}


\date{\today}

\begin{abstract}
We propose a novel
method to measure the 
neutron-neutron scattering length using the \(^{6}\mathrm{He}(p,p\alpha)\)$nn$ reaction in inverse kinematics at high energies.
The method is based on the final state interaction (FSI) between the neutrons after the sudden knockout of the $\alpha$ particle. 
We show that the details of the neutron-neutron relative energy distribution allow for a precise extraction of the $s$-wave scattering length.
We present the state-of-the-art in regard to the theory of this distribution. The distribution is calculated in two steps. 
First, we calculate the ground-state wave function of \(^6\)He as a \(\alpha n n\) three-body system.
For this purpose we use Halo effective field theory (Halo EFT), which also provides uncertainty estimates for the results.
We compare our results at this stage to model calculations done with the computer code FaCE.
In a second step we determine the effects of the \(nn\) FSI using the \(nn\) t-matrix.
We compare these FSI results to approximate FSI approaches based on standard
FSI enhancement factors.
While the final distribution is sensitive to the \(nn\) scattering length, it depends only weakly on the
effective range. Throughout we emphasize the impact of theoretical uncertainties on the neutron-neutron relative energy distribution, and discuss the extent
to which those uncertainties limit the extraction of the neutron-neutron scattering length from the reaction  \(^{6}\mathrm{He}(p,p\alpha)\)$nn$. 
\end{abstract}

\maketitle

\section{Introduction and Conclusion}
The significant difference between the proton-proton (\(pp\)) and the neutron-neutron (\(nn\)) interaction 
is a consequence of the charge symmetry breaking of the nucleon-nucleon (\(NN\)) interaction. It
has its fundamental origin in the different masses and electromagnetic properties of the 
light quarks \cite{Mill06}. The charge symmetry breaking of the \(NN\) interaction is, for example, 
manifested in the  $s$-wave scattering lengths that parameterize the zero-energy \(NN\) cross section.
Because of their fundamental importance, the \(nn\) and \(pp\) scattering lengths have been a topic of intense research.
The current accepted values are $a_{pp}^\mathrm{str}=(-17.3 \pm 0.4)$ fm
and  $a_{nn}^\mathrm{str}=(-18.9 \pm 0.4)$ fm \cite{Gard09,Mach01,GaPh06}. The superscript ``str" indicates that electromagnetic effects have been removed
in these numbers, but
in the remainder of the paper we actually use the raw quantities measured in experiment. The corresponding value for the \(nn\) interaction is
\(a_{nn}=(-18.6 \pm 0.4)\)~fm.

It should be noted, however, that there is a systematic and significant difference between the extracted values of $a_{nn}$ from
neutron-induced deuteron breakup reactions measured by two different collaborations with different experimental setups.
A group from Bonn has measured the \(d(n,pn)n\) reaction and extracted $a_{nn}=-16.3(4)$ fm \cite{Huhn00}
using a theoretical analysis based on three-body Faddeev equations \cite{Mach01a}.
Different beam energies and analysis methods (absolute vs. relative cross sections) yielded slightly different, but consistent, values for the scattering length.
This result was confirmed (but with larger uncertainties) in a more recent measurement in Bonn using the same reaction but with only the final state proton being
detected \cite{Wits06}.
Around the same time as the earlier Bonn experiment, a group from TUNL extracted the value $a_{nn}=-18.7(7)$ fm \cite{GonzalezTrotter:1999zz}\footnote{
  The value is from Ref. \cite{GonzalezTrotter:1999zz}, but the uncertainty band is from Ref. \cite{Gonz06}, 
  where almost the same group published a reanalysis of the data.
}
from their experiment using the same reaction with all final particles detected
and the same theoretical treatment.
This value was later confirmed in a reanalysis of the TUNL experiment \cite{Gonz06}.
The discrepancy between the two values is an unsolved puzzle and points towards an unknown experimental systematic uncertainty.

An alternative method which avoids the complication of the hadronic three-body final state is given by
the pion capture reaction \(\pi^- d \rightarrow nn\gamma\). In this case, a slow pion is captured in
a \(^2\mathrm{H}\) atomic state and then absorbed by the deuteron yielding the breakup into two neutrons and a photon.
In some experiments only the high-energy photon is measured, in others the photon is measured in coincidence with one
of the outgoing neutrons. The scattering length is extracted from a fit to the shape of the neutron
spectrum, i.e., the decisive feature is the relative height of the quasi-free \(p\pi\) capture peak and the 
peak caused by final-state interaction (FSI). From the combination of experiments at 
PSI \cite{Gabioud:1979uz,Gabioud:1981sw,Gabioud:1984fs,Scho87} and Los Alamos \cite{Howe98} \(a_{nn} = -18.6(4)\)~fm is deduced
\cite{Mach01}, which is in agreement with
the deuteron breakup experiments at TUNL and presently considered the
accepted value. 

\begin{figure}[htb!]
  \centerline{\includegraphics[width=0.7\textwidth]{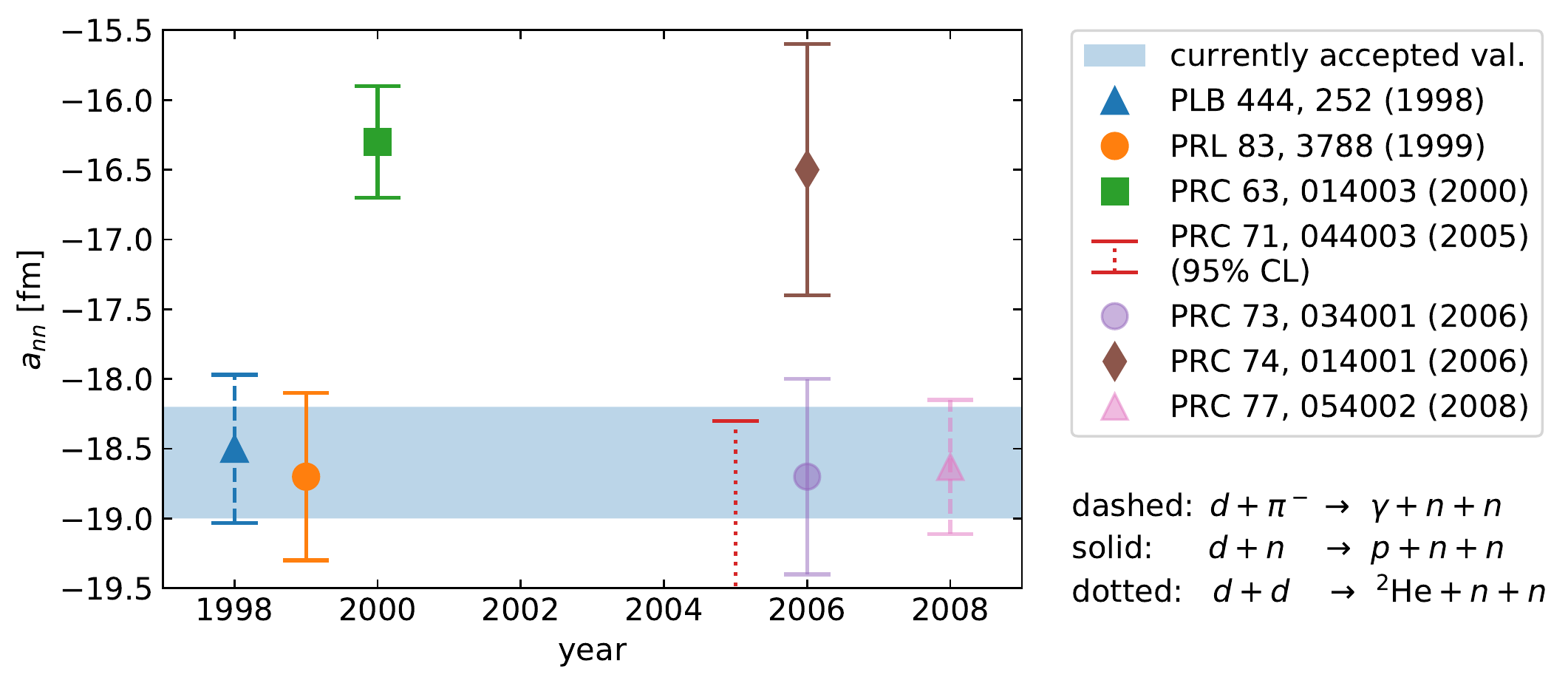}}
  \caption{
    Recent experimental data on the neutron-neutron scattering length, as reported in 
  Refs. \cite{Howe98,GonzalezTrotter:1999zz,Huhn:2001yk,Huhn00,Gonz06,Wits06,Chen:2008zzj,Baumer:2005kr}. 
   The horizontal band displays the uncertainty band of the accepted value, according to Ref. \cite{Mach01} and Ref. \cite{Gard09}. 
  The line style of the error bar encodes information on the reaction, cf. legend.
  Results based on the same experimental data only differing in the analysis use the same point style.
  In the shown values for \(a_{nn}\), effects of the magnetic-moment interaction
  are not removed,
  see, e.g., Ref. \cite{Gard09} for more details.} \label{fig0}
\end{figure}

The most recent data for \(a_{nn}\) are displayed in \cref{fig0} together with
the limits of the accepted value (horizontal band). We also display there a result obtained at KVI from  
the reaction \(^2\mathrm{H}(d,^2\mathrm{He})^2n\).
Using a simple reaction model, they extracted an upper bound of -18.3 fm
at the 95\% confidence level.

The \(nn\) scattering length can be also inferred from pion-photoproduction on deuterium \(\gamma d \to \pi^+ nn\) as, e.g., Refs. \cite{Tzara:1976yy,Faldt:1986su,Lensky:2007zc,Nakamura:2020qkz} have shown.
While the theoretical study in Ref. \cite{Lensky:2007zc} used chiral perturbation theory for the regime of \(\gamma\)
energies close to pion-photoproduction threshold, it is supplemented for higher \(\gamma\) energies by a recent study in Ref. \cite{Nakamura:2020qkz} using a realistic model for this reaction. 
The determination of the \(nn\) scattering length with such an experiment can be realized by a precision measurement of the energies of the incoming \(\gamma\) and the outgoing \(\pi^+\).
The neutron detection efficiency is not problematic for this experimental method
but an analysis of the theory uncertainties remains to be carried out.

In this article we propose a novel method to measure  \(a_{nn}\). This method
takes advantage of inverse kinematics at a \hesix\ beam energy of a few hundred
MeV/nucleon. The \hesix\ beam impinges on a proton target, resulting in
quasi-free knockout of the \(\alpha\) particle.
The two halo neutrons of the \hesix\ projectile are
liberated by this knockout, and both continue flying forward in the laboratory
system with approximately beam velocity. Their relative energy remains small: it
is determined by the overlap of the \(nn\) wave functions in the \hesix\ ground-state and
the \(nn\) scattering state, and so depends strongly on the \(nn\) scattering length at
low relative energies. The neutrons are detected at approximately 10 m distance
from the target around zero degrees with a \(1 \times 1~\textrm{m}^2\) large detector array covering
the \(nn\) relative-energy spectrum from 0 to 1 MeV. 
In addition to the two halo
neutrons, both the \(\alpha\) particle and proton are detected, allowing selection of
events in which the charged particles are scattered to large angles. This
quasi-elastic high-energy scattering process results in large relative energies
between the charged particles and the neutrons: that this occurs is verified by
the measurement of the tracks of the two scattered particles. Any non-\(nn\)
final-state interactions are now again high-energy scattering processes,
resulting in substantial changes of angles and relative energies between the
particles. In particular, the neutrons will not remain in the low-relative-energy state
(\(E_{nn}<1\) MeV) in the case of final-state interaction with the charged particles,
and so will not be detected—and the kinematical signature of energies and angles
of charged particles would also then not correspond to the quasi-elastic
kinematics. Therefore, although the non-\(nn\) FSI is present, it results only in a
reduction of observed events and does not distort the low-energy \(E_{nn}\) spectrum.
Nowadays the relative energy between the high-energy neutrons can be
measured with an energy resolution of about 20 keV, so the \(nn\) energy
spectrum can be mapped out with high accuracy. In this work we argue that the
imprint of \(a_{nn}\) on that low-energy \(nn\) (relative) energy spectrum far
exceeds uncertainties coming from the \(^{6}\mathrm{He}\)  structure and the
reaction dynamics. And, because all four
final-state particles are detected, a background-free measurement can be
performed. We conclude that this kind of
\(^{6}\mathrm{He}(p,p'\alpha)nn\) 
measurement can be used to extract the neutron-neutron scattering length. A proposal
to carry out such an experiment has been approved at RIKEN \cite{nn_scat_len_ribf_prop2018}.

In contrast to the novel method proposed in this work, previous extractions of  $a_{nn}$ relied on measuring the intensity of the neutron-neutron 
final state interaction (FSI) peak relative to the quasi-free peak. These 
are located at rather different neutron energies. 
In our proposal, the neutrons originating from the projectile have a high and almost constant
velocity avoiding possible sources of systematic uncertainties due to energy-dependent corrections for neutron 
efficiency, scattering and attenuation in the target etc., as were necessary in the previous experiments.

It is clear that a general description of the \(^{6}\mathrm{He}(p,p\alpha)2n\) knockout reaction is a formidable task. However, we stress that the proposal is to extract $a_{nn}$ in high-energy kinematics where
we argue that
the $a_{nn}$-dependence of the relative-energy distribution of the neutrons can be calculated reliably and with quantified theoretical
uncertainties.
An analogous extraction could be made in the \(t(p,2p)2n\) knockout reaction as a cross-check, but
we will not consider that reaction in this paper.

The calculations presented below demonstrate the sensitivity of the \(nn\) relative-energy distribution in  \(^{6}\mathrm{He}(p,p\alpha)2n\) to the \(nn\) scattering length. In \cref{sec:nn_scat_length} we show that varying the nominal scattering length of \SI{-18.7}{\fm} by \SI{2}{\fm} changes characteristic parts of the
distribution around \(E_{nn}=\SI{100}{\kev}\) by roughly 10\%.
This sensitivity will enable a precise experimental determination of the scattering length.
Furthermore, the spectrum has only a small
dependence on the \(nn\) effective range: we find that its effect is 
less than 1\%. 
We now lay out the procedure and assumptions through which we calculate the
\(nn\) relative-energy distribution. We also summarize already here the
uncertainties associated with each piece of our calculation. 
\begin{enumerate}
\item In \cref{sec:tb_calcs} and \cref{sec:nn_gs_distrib} we describe and
present results from our computation of the ground-state momentum distribution
of \hesix, i.e., the distribution in the absence of final-state interactions.
We treat the nucleus as a \(\alpha n n\) three-body system and use both an EFT
of the halo nucleus~\cite{Ji:2014wta,Gobel:2019jba} and a three-body model. The
EFT calculation is carried out at leading order and has a nominal uncertainty of
$\approx 20$\% for $E_{nn} \approx 1$ MeV.  We compare the EFT momentum
distribution to that obtained with a three-body model of \hesix\ that uses local
Gaussian two-body potentials as well as a three-body force. This  ``\lgmacronym"
calculation is in the tradition of, e.g.,
Refs.~\cite{Chulkov:1990ac,Zhukov:1993aw}, and is implemented  via the computer
code FaCE \cite{Thompson:2004dc}. The resulting momentum distribution is
consistent with that obtained from the Halo EFT within the expected uncertainty
of a leading-order calculation. At next-to-leading order the EFT uncertainty
band will be smaller and better agreement is expected. We analyze which effects
cause the differences between the EFT and \lgmacronym~distribution and show that
corrections to the \(nc\) t-matrix are the most significant NLO corrections to
the structure of \hesix~in the EFT approach.

\item In treating the reaction dynamics we assume the knockout of the \(\alpha\)
by the proton results in its sudden removal and does not distort the \(E_{nn}\)
spectrum. To some degree this can be ensured during the analysis of the
experimental data by taking only those events where the measured charged
particles meet the necessary kinematical conditions. Nevertheless, assessing the
error induced in the neutron-neutron relative-energy spectrum by this use of the
sudden approximation is an important topic for future work.

 \item At the moment only \(nn\) FSI is taken into account. 
  The reason that the distortion
  effect due to $\alpha n$  or $pn$ FSI is higher order is the
  smallness of the effect in the chosen kinematics according to
  scaling arguments. The $\alpha n$ FSI is suppressed by the ratio
  $p_{nn}/k$, where $p_{nn}$ and $k$ are the relative momentum between
  the two neutrons and the momentum transfer to the $\alpha$ particle,
  respectively. The experiment in the proposed kinematics selects by
  construction small $p_{nn}$ and large $k$, resulting in the suppression
  of effects due to FSI between the neutrons with the charged particles
  involved in the reaction. Any remaining correction, if necessary, will
  be small, so that the accuracy of calculating the correction does not have
  to be high to fulfil the precision requirement of the analysis of experiment
  and the extraction of $a_{nn}$.

\item Under the assumptions of the previous two points the \(nn\)
 relative-energy distribution is straightforwardly obtained from the \hesix\
 wave function using a two-body treatment of FSI. In \cref{sec:final_nn_distrib}
 we compare results based on a full calculation of the \(nn\) FSI using the
 t-matrix to approximate results based on so-called enhancement factors. We
 discuss the derivation of the latter technique, which was established by Watson
 and Migdal \cite{Watson:1952ji,Migdal:1955_1}. (Reviews can be found in Refs.
 \cite{Slobodrian:1971an,goldberger2004collision}.)
By formulating the problem in terms of two-potential scattering theory we show
that the enhancement factors are approximations to an exact calculation of
\(nn\) FSI via the t-matrix.  
This establishes a preference for the t-matrix approach. But, regardless of that
preference, the calculations of \cref{sec:final_nn_distrib} show that the key
parts of the distribution around \(E_{nn}=\SI{100}{\kev}\) that change by
roughly 10\% if \(a_{nn}\) is varied by 2 fm (cf. \cref{sec:nn_scat_length}) are
moderately insensitive to the approach used for the \(nn\) final-state
interaction. We conclude the impact of different treatments of \(nn\) FSI on the
error budget of the \(a_{nn}\) extraction is minimal. 

\item Our calculations of the neutron energy distribution after $\alpha$-particle knockout use only the partial-wave state where the \(nn\) system
and the \((nn)-c\) system are both in a relative \(s\)-wave. This is the most important
component in our calculation of the ground state, and
 the \(nn\) FSI increases its dominance. We quantitatively
assessed the relevance of the other components and found that in the case of a LGM
calculation with \(nn\) FSI the contribution to the \(nn\) energy distribution for \(E_{nn} < 1\) MeV
from two neutrons in a relative \(^3P_1\) wave
is at least a factor of 30 smaller than that
that from the \(^1S_0\) wave.
\end{enumerate}
Having laid out the procedure for calculating the \(nn\) relative-energy
distribution in Sections \ref{sec:tb_calcs}--\ref{sec:final_nn_distrib} in
\cref{sec:nn_scat_length} we investigate the sensitivity of that distribution to
the scattering length. We close 
in \cref{sec:outlook} with an outlook regarding future calculations.

\section{\texorpdfstring{Three-body calculations of $\boldsymbol{^6}$He}{Three-body calculations of He-6}}\label{sec:tb_calcs}
The first step for obtaining the \(nn\) relative-energy distribution is obtaining the ground-state wave
function of \(^6\)He. Because of its halo structure it can be described as a \(\alpha n n\) three-body system.
The halo structure manifests itself in a two-neutron separation energy, which corresponds to the binding energy of the \(\alpha n n\) three-body
system, \(\btz = \SI{0.975}{\mev}\) \cite{Brodeur:2012zz}, much smaller than the \(\alpha\) core's excitation energy \(E_\alpha^* \approx \SI{20}{\mev}\).
We calculate the wave function in Halo EFT at leading order and compare to results obtained in three-body
model calculations of the system.
In this section, we introduce concepts and quantities necessary for both methods.
The relative positions and momenta can be described by splitting the three-body system into a two-body system and a third particle.
The momenta are then given in terms of the relative momentum between the constituents of the subsystem and the relative momentum between the 
third particle and the center of mass of the subsystem. In position
space, the coordinates can be chosen analogously.
The third particle is called spectator, its choice is arbitrary.
These momenta are called Jacobi momenta.
The Jacobi momenta of the three-body system with momenta \(k_i\) and masses \(m_i\) (\(i \in \{1,2,3\}\)) are given by
\begin{align}
  \v{p}_i &\coloneqq \mu_{jk}      \K{ \frac{\v{k_j}}{m_j} - \frac{\v{k_k}}{m_k}              } \,, &
  \v{q}_i &\coloneqq \mu_{i\K{jk}} \K{ \frac{\v{k_i}}{m_i} - \frac{\v{k_j} + \v{k_k}}{M_{jk}} } \,, 
\end{align}
where the definitions \(\mu_{ij} \coloneqq \K{m_i m_j}/\K{m_i + m_j}\), 
\(\mu_{i\K{jk}} \coloneqq \K{m_i M_{jk}}/\K{m_i + M_{jk}}\) and \( M_{ij} \coloneqq m_i + m_j \) hold.\footnote{
    Note, that, e.g., in Ref. \cite{Zhukov:1993aw} a different convention for the Jacobi momenta is used.
    Some notes on the differences can be found in the supplemental material \cite{supp_mat}.
    We use the convention which is used, e.g., in Refs. \cite{Gobel:2019jba,Hammer:2017tjm,Ji:2014wta}.
}
In order to describe this three-body system in a partial-wave basis we have to assign quantum numbers.
With the coordinates, they generally depend on the chosen spectator.
The relative orbital angular momentum quantum number of the subsystem is given by \(l\),
the one between the third particle and the subsystem is given by \(\lambda\).
The quantum number \(s\) specifies the total spin of the subsystem, while \(\sigma\) denotes the spin
of the third particle. In \jJcoupling\ the relations \(\v{j}=\v{l}+\v{s}\) and \(\v{I}=\v{\lambda}+\v{\sigma}\)
hold.
A general partial wave state reads
\begin{equation}
    \iket{\K{l,s}j, \K{\lambda, \sigma} I;J,M}{i}\,,
\end{equation}
where the index \(i\) on the right specifies the spectator.

Before going into the specifics of the three-body calculations, we want to discuss the
reference states for calculating the ground-state wave function.
Since we are investigating ground states, we have in the case of \hesix~the condition \(J=M=0\).
The spin of the \(\alpha\) particle is zero.
This implies \(\sigma=0\) in case of the \(\alpha\) particle as spectator (indicated by an index \(c\) for core at the bra/ket).
The two neutrons with spin \(1/2\) can couple to \(0\) or \(1\), meaning that \(s\) is \(0\) or \(1\).
Under these conditions four different types of partial-wave basis states can be formed,
states of each type are parameterized by the orbital angular momentum quantum number of the subsystem \(l\).
If \(s=0\), the states are of the form
\begin{equation}
  \iket{\Omega_c^{(0,l,l)}}{c} \coloneqq \iket{\K{l,0}l, \K{l, 0} l;0,0}{c} \quad \textrm{with~} l \geq 0 \,.
\end{equation}
For \(s=1\) the following three types of states can be formed:
\begin{alignat}{2}
  \iket{\Omega_c^{(1,l,l-1)}}{c}  &\coloneqq \iket{\K{l,1}l-1, \K{l-1, 0} l-1; 0,0}{c} && \quad \textrm{with~} l \geq 1 \,, \\
  \iket{\Omega_c^{(1,l,l)}}{c}    &\coloneqq \iket{\K{l,1}l,   \K{l,   0} l;   0,0}{c} && \quad \textrm{with~} l \geq 1 \,, \\
  \iket{\Omega_c^{(1,l,l+1)}}{c}  &\coloneqq \iket{\K{l,1}l+1, \K{l+1, 0} l+1; 0,0}{c} && \quad \textrm{with~} l \geq 0 \,.
\end{alignat}
This produces a complete, orthogonal angular-momentum 
basis for a three-body system of \(J=M=0\) that is formed out of two distinguishable spin-\(\oh\) particles and one spin-0 particle.
In the following we will call these basis states ``reference states" and calculate their overlaps with the
eigenstate of the three-body Hamiltonian in order to obtain wave functions on the partial-wave basis.

  The ground state of \hesix~has positive parity and is antisymmetric under interchange of the two neutrons.
  Only the piece of a reference state with the same symmetries as the ground state will have non-vanishing overlap with it.
  Therefore, \(\Omega_c^{(1,l,l-1)}\) and \(\Omega_c^{(1,l,l+1)}\) are not suitable reference states, as they have negative parity.
  Similarly, the requirement of \(nn\)-antisymmetry means that we only need to consider states \(\Omega_c^{(0,l,l)}\) where the quantum number \(l\) is even,
  together with states \(\Omega_c^{(1,l,l)}\) where \(l\) is odd.
This analysis of the possible states is consistent with other three-body calculations of the ground state of \hesix~presented, e.g.,
in Refs. \cite{Chulkov:1990ac,Zhukov:1993aw}.

  \subsection{\lgmnameC}
    Our model calculation of the three-body system employs commonly used local \(l\)-dependent Gaussian potentials
    as well as a three-body force.
    We call this a \lgmname~(\lgmacronym). 
    To solve it for the three-body system we use the computer code FaCE \cite{Thompson:2004dc}\footnote{
      The code itself can be obtained from a research data repository
      (DOI: \href{https://doi.org/10.17632/4g97cjzzyp.1}{10.17632/4g97cjzzyp.1}).
    }.
    It calculates the position-space wave function of a three-body system by solving the Schrödinger equation
    with local \(l\)-dependent two-body interactions and phenomenological three-body potentials.
    It is capable of removing unphysical bound-states from two-body potentials via the supersymmetric (SUSY) transformations
    described in Ref. \cite{Sparenberg:1997}.
    The name FaCE is an acronym for ``Faddeev with Core Excitation".
    It alludes to the fact that in the default setting it solves not the Schrödinger equation but the equivalent
    Faddeev equations.\footnote{
      Note, that the Faddeev equations which are used by FaCE are equivalent to the ones used
      in our EFT calculation. However they are not of the same form.
      FaCE uses matrix elements of potentials, the EFT calculation uses matrix elements of t-matrices.
      These two versions of the Faddeev equations have the decomposition of the total state into
      components in common.}
    As the name also expresses core excitation effects can be taken into account within this code.

    This code and its ancestors were used for calculations of multiple nuclei.
    In the case of \hesix, the position-space probability densities and transverse-momentum distributions
    were already calculated with ancestors of the FaCE code, e.g., in Refs. \cite{Chulkov:1990ac,Zhukov:1993aw}.
    The results for the transverse momentum distribution agree well with available experimental data.
    
    We now define the parameters that are specified in a typical FaCE input file for \hesix, and in the process also write down
    the potentials employed in our \lgmacronym~for this system. 
    For the \(nn\) interaction as well as the \(n\alpha\) interaction we use local central and spin-orbit potentials:
    \begin{align}
      \mel{r;l,s}{ V_c^{(\tilde{l})} }{r^\prime;l^\prime,s^\prime}   &\coloneqq \kd{l}{l^\prime} \kd{l}{\tilde{l}} \kd{s}{s'} \frac{\de{r^\prime-r}}{r'^2}
        \bar{V}_c^{(l)} \exp\K{-r^2/ \K{a_{c;l }^2 }}                      \,,  \label{eq:V_c} \\
      \mel{r;l,s}{ V_{SO}^{(\tilde{l})} }{r^\prime;l^\prime,s^\prime} &\coloneqq \kd{l}{\tilde{l}} \frac{\de{r^\prime-r}}{r'^2}
        \bar{V}_{SO}^{(l)} \mel{l,s}{\v{L}\v{S}}{l^\prime,s^\prime} \exp\K{-r^2/ \K{a_{SO;l}^2 }} \,,  \label{eq:V_so}
    \end{align}
    where the depth parameters are denoted by \(\bar{V}_c^{(l)}\) and \(\bar{V}_{SO}^{(l)}\).
    The range parameters are given by \(a_{c;l }\) and \(a_{SO;l}\).

    In the \lgmacronym~calculation the \(nc\) interaction is present in the \swave, \pwave~and \dwave.
    In the \pwave~and \dwave~both central potentials and spin-orbit potentials are used.
    A \(p\)-wave nucleon-$\alpha$ potential of this form was first specified in Ref. \cite{Sack:1954zz}, where it was shown to provide a reasonable description of low-energy \(p\alpha\) phase shifts. 
    Here we take for all \(nc\) potentials a range of \(a_{c;l} = a_{SO;l} = \SI{2.3}{\fm}\).
    The depth parameters are: \(\bar{V}_{c;0} = \bar{V}_{c;1} = \SI{-47.32}{\mev}\), \(\bar{V}_{SO;1} = \bar{V}_{SO;2} = \SI{-11.71}{\mev}\),
    and \(\bar{V}_{c;2} = \SI{-23.0}{\mev}\).
    These parameters except the \swave~ones were inter alia used for the calculation with an ancestor of FaCe in Ref. \cite{Zhukov:1993aw} and presumably in Ref. \cite{Chulkov:1990ac} as well as in the recent \cite{Grigorenko:2020jts}.
    In the case of the \swave~we do not use the repulsive potential used in Ref. \cite{Zhukov:1993aw}. Instead, 
    we follow the FaCE sample input file for \hesix~which means we use the previously mentioned attractive potential and remove the
    unphysical \(n\alpha\) bound state using the SUSY transform capabilities of FaCE.
    This attractive potential produces a satisfactory fit to the phase shifts given in Ref. \cite{Ali:1984ds}.

    For the \(nn\) interaction we use a \swave~central potential with the parameters \(\bar{V}_{c;0} = \SI{-31.0}{\mev}\) and
    \(a_{c;0} = \SI{1.8}{\fm}\).
    These parameters were also used inter alia for the calculation in Ref. \cite{Zhukov:1993aw,Grigorenko:2020jts} and are taken from Ref. \cite{Brown:1976}.

    The phenomenological three-body force reads
    \begin{equation}
      V_\tbi{\K{\rho}} \coloneqq \frac{s_\tbi}{ 1.0+\K{\rho/\rho_\tbi}^{a_\tbi}} \,.
    \end{equation}
    The parameters \(\rho_\tbi = \SI{5.0}{\fm}\) and \(a_\tbi = 3\) are used, as they are set in the sample
    input file of FaCE.
    The depth parameter \(s_\tbi\) will be tuned to reproduce \(\btz\).

    FaCE calculates the wave functions of three-body systems such as \hesix~in terms of a decomposition in the
    hyperangular momentum \(K\).
    The single components are specified by the hyperangular momentum quantum number
    and the angular and spin quantum numbers.
    By doing the decomposition in \(K\), the wave function's coordinate-space dependence on \(x\) and \(y\), which are
    conjugate to \(p\) and \(q\),
    can be replaced by the dependence on the hyperradius \(\rho \coloneqq \sqrt{x^2 + y^2}\).
    In the following these wave function components \(\chi_K{\K{\rho}}\) will have only the additional indices \(l\) and \(S\),
    as, due to the symmetries discussed in the beginning of this section, these determine all other quantum numbers of the \hesix~ground state with \(J=M=0\) and positive parity
    in \jJcoupling: \(\lambda=l\), \(s=S\), \(\sigma=0\) and \(j=I=l\) (core as spectator).
    The \(nn\) relative-momentum distribution is calculated using the momentum-space wave function, while
    the calculation of this wave function from the \(\cf{K,l}{S}{\rho}\) is summarized in the supplemental material \cite{supp_mat}.
    It contains also details on the computational parameters of the model calculation.

  \subsection{Halo EFT approach}
    A second approach for obtaining the three-body wave function of \hesix~is using Halo effective field theory (Halo EFT).
    An effective field theory is a toolkit for exploiting the scale separation of a physical system 
    in order to calculate
    observables as a series in the ratio
    of a typical momentum scale over a high momentum scale. The high momentum scale is the lowest scale of omitted physics.
    Systematic improvement of the results is then possible by calculating higher orders in the expansion. And at any given order
    the EFT's expansion in a ratio of momentum scales enables robust uncertainty estimates for its predictions.
    
    Halo EFT is a pionless EFT describing halo nuclei.
    The halo nucleons are associated with the lower momentum scale while the high momentum is associated with effects such as pion creation, removal of nucleons
    from the core or excitation of the core.
    In the case of \hesix~the low-momentum scale \(\mlow\) can be determined, using the three-body binding energy \(\btz=\SI{0.975}{\mev}\),
    to be \(\mlow  = \sqrt{m_n \btz} \approx \SI{30}{\mev}\).
    The high-momentum scale is given by \(\mhigh = \sqrt{m_n E_\alpha^*} \approx \SI{140}{\mev}\),
    where the excitation energy of the core is given by \(E_\alpha^* \approx \SI{20}{\mev}\).
    The basic ingredient of an EFT calculation is the power counting.
    It tells which terms are of which order in \(\mlo/\mhigh\) and thereby defines which have to be included in a calculation at a given order.
    The \(n\alpha\) system was first investigated in a Halo EFT framework in Refs. \cite{Bertulani:2002sz} and \cite{Bedaque:2003wa},
    which proposed different power countings: Ref. \cite{Bertulani:2002sz} proposes \(a_1 \sim \mlow^{-3}\)
    and \(r_1 \sim \mlow\), where  \(a_1\) is the \pwave~scattering volume and \(r_1\) the \pwave~effective range.
    Usually one expects that the effective range parameters are of order of the appropriate power of \(\mhigh\), thereby we
    have two fine-tunings here.
    According to the power counting of Ref. \cite{Bedaque:2003wa} \(a_1 \sim \mlow^{-2} \mhigh^{-1}\)
    and \(r_1 \sim \mhigh\) hold. This power counting has
    the minimum number of fine-tunings necessary to produce a bound state or resonance in the low-energy region of the EFT.

    The latter power counting was used in Ref. \cite{Ji:2014wta}, where Halo EFT was applied to \hesix.
    In that paper the two-body subsystems as well as the three-body system were successfully renormalized.
    In order to renormalize the three-body system with a three-body force the binding energy \(\btz\) was used as input.
    Additionally, Faddeev amplitudes were calculated and their independence of sufficiently high cutoffs was demonstrated.
    This work was continued in Ref. \cite{Gobel:2019jba}, where ground state probability densities were calculated in Halo
    EFT.
    The potentials corresponding to the leading-order t-matrices used in Refs. \cite{Ji:2014wta,Gobel:2019jba}
    are energy-dependent. 
    While in the case of the \swave~\(nn\) interaction this dependence vanishes in the limit that the cutoff goes to infinity, in the case
    of the \pwave~\(nc\) interaction it does not vanish.
    In Refs. \cite{Mckellar:1984zq,formanek04} quantum mechanics with energy-dependent potentials is discussed.
    Inter alia a modified normalization condition for wave functions is derived. 
    These findings were applied to the calculation of the probability density in Ref. \cite{Gobel:2019jba}, where
    it was found that these modifications are negligible in the low-energy region. 
    Furthermore, the robustness of the results with respect to the regulator was checked.
    The probability density is independent of the cutoff and the form of the momentum-space regulator.

    In this paper, we use Halo EFT to calculate the ground state wave function of \hesix.
    We base our calculation on the methodology used in Ref. \cite{Gobel:2019jba}.
    We will solve the same Faddeev equations as in Ref. \cite{Gobel:2019jba}, the only difference is that
    we will not calculate overlaps of plane wave states with \(\ket{\Psi}\) but overlaps of partial wave states
    with \(\ket{\Psi}\).
    Since partial wave states were also widely used in that paper, many formulas can be reused.
    At this point, we briefly review the Faddeev equations.
    While they can be derived in a non-relativistic (effective) field theory, see, e.g., Ref. \cite{Hammer:2017tjm},
    we describe here the connection to the Schrödinger equation.
    This allows for straightforward comparisons with quantum mechanical model calculations.
    The Schrödinger equation for a three-body system with a kinetic Hamilton operator \(H_0\), two-body interactions \(V_i\)
    and a three-body potential \(V_3\) reads
    \begin{equation}
      \K{ H_0 + \sum_i V_i + V_3 } \ket{\Psi} = E_3 \ket{\Psi} \,,      
    \end{equation}
    where \(E_3\) is the energy of the three-body system.
    The index \(i\) is the index of a third particle defining the subsystem consisting of the remaining particles, in which \(V_i\) acts.
    Accordingly \(i \in \{c,n,\np\}\) holds.
    For the moment we consider the system without the three-body force.
    The Schrödinger equation can be rewritten into a set of coupled equations,
    the so-called Faddeev equations, for the
    Faddeev amplitudes \(\ket{F_i}\), see, e.g., Refs. \cite{Afnan:1977pi,gloeckle83,Ji:2014wta}:
    \begin{equation}
      \ket{F_i} = \sum_{j \neq i} G_0 t_j \ket{F_j} \,, \label{eq:f_i}
    \end{equation}
    where the connection to the desired \(\ket{\Psi}\) is given by
    \begin{align}
      G_0 t_i \ket{F_i} &= G_0 V_i \ket{\Psi} \,, \\
      \sum_i G_0 t_i \ket{F_i} &= \ket{\Psi} \,. \label{eq:psi_f_i}
    \end{align}
    We will use the latter equation in order to obtain the ground-state wave function from
    the Faddeev amplitudes.
    When we solve the Faddeev equations numerically, we have to use a representation of the states.
    It is common to use the following representation for the \(\ket{F_i}\):
    \begin{equation}
      F_i{\K{q}} = \rint{\p} \reg{i}{p} \ibraket{i}{p,q;\Omega_i}{F_i}{} \,, \label{eq:rep_f_i}
    \end{equation}
    where we assumed that \(V_i\) acts only in one partial wave channel given by the
    set of quantum numbers \(\Omega_i\) seen from particle \(i\) and that
    it has a one-term separable form in the corresponding two-body subsystem.
    
    The three-body force can be included in the Faddeev formalism in several
    ways.
    One way is to modify \cref{eq:f_i} and leave
    the relation of obtaining the full states from the Faddeev amplitudes, namely \cref{eq:psi_f_i},
    unchanged~\cite{gloeckle83}.
    We use this method, the employed three-body force is given in Ref. \cite{Ji:2014wta}.
    Alternative possibilities for this force in the case of \hesix~can be found in Ref. \cite{Ryberg:2017tpv}.

    We now give some more details on the used two-body interactions.
    Since we solve Faddeev equations in momentum space, which are equivalent to the Schrödinger equation,
    the two-body interactions are specified in the form of t-matrices.
    The real part of the denominator of the t-matrix corresponds to an effective-range expansion, which is carried out up to a certain order that is
    determined by the power counting of the EFT and the order of the calculation. 
    But this does not determine the (off-shell) t-matrix.
    For convenience in the implementation of the Faddeev equations, we use separable t-matrices corresponding to
    separable potentials.
    This is a common choice, see, e.g., Refs. \cite{Ji:2014wta,Gobel:2019jba}.
    The elements of the t-matrix describing the interaction between particles \(i\) and \(j\) read
    \begin{equation}\label{eq:t_2bd}
      \mel{p,l}{t_{ij}{\K{E}}}{\pp,l^\prime} = 4\pi \kd{l}{l^\prime} \kd{l_{ij}}{l} \reg{{ij}}{p} \tau_{ij}{\K{E}} \reg{{ij}}{\pp} \,,
    \end{equation}
    where \(l_{ij}\) specifies the quantum number \(l\) of the interaction channel.
    The functions \(\reg{{}}{p}\) are regulator functions specifying the damping at and above momenta
    of the order of the cutoff scale \(\beta\) parameterizing these functions.
    Additionally, they determine the off-shell behavior of the t-matrices.
    We use \(\reg{{}}{p} = p^l \theta{\K{\beta-p}}\).
    For our three-body calculation we have to embed the t-matrix into the three-body system and take matrix elements
    of this embedded version. We obtain for the elements of the matrix \(t_i\) describing the interaction given by spectator \(i\), i.e. the one between \(j\) and \(k\), the following expression:
    \begin{equation}\label{eq:t_3bd}
      \imel{i}{p,q;\Omega}{t_i{\K{E_3}}}{\pp,\qp;\Omega^\prime}{i} = \kd{\Omega}{\Omega^\prime} \kd{\Omega}{\Omega_i} 
      \mel{p,l_i}{t_{jk}{\K{ E_3- \frac{q^2}{2 \mu_{i\K{jk}} } }}}{\pp,l_i} \,,
    \end{equation}
    where \(l_i=l{\K{\Omega_i}}\) is the subsystem orbital angular-momentum quantum number of the interaction channel
    given by the multiindex \(\Omega_i\).
    The reduced t-matrix elements \(\tau_{jk}{\K{E}}\) contain the first terms of the effective range expansion in their denominators.
    In our leading-order Halo EFT for \hesix, they are given by
    \begin{align}\label{eq:tau_nn}
      \tau_{nn}{\K{E}} &= \frac{1}{4\pi^2 \mu_{nn}} \frac{1}{\gamma_0 + \ci k} \,,\\
      \tau_{nc}{\K{E}} &= \frac{1}{4\pi^2 \mu_{nc}} \frac{1}{\gamma_1 \K{k^2 - k_R^2}} \,, \label{eq:tau_nc}
    \end{align}
    whereby the relation \(k = \sqrt{2\mu_{jk}E}\) holds.
    The parameter \(\gamma_0\) is the momentum of the \(nn\) virtual state and at leading order is given by
    the \(nn\) scattering length via \( \gamma_0 = a_{0}^{-1} \).
    In contrast to the \(nn\) interaction, the \pwave~$\tau_{nc}$ does not contain a unitarity term at leading order according to the power
    counting.
    The \(nc\) interaction is parameterized by the effective range expansion parameters \(a_1\) and \(r_1\) via
    \(\gamma_1 = -r_1/2\) and \(k_R = \sqrt{2/\K{a_1 r_1}}\), whereby \(k_R\) is the momentum of the low-energy resonance.
    The values \(r_1 = -174.0227\)~MeV and \(k_R = 37.4533\)~MeV were used.
    They can be obtained from the \(a_1\) and \(r_1\) given in Ref. \cite{Arndt:1973ssf}.
    The core mass is approximated by \(m_c \approx 4 m_n\).

    After discussing the interactions, we briefly describe how the wave functions are obtained.
    From the system of equations for the Faddeev amplitudes given in \cref{eq:f_i} one obtains
    a coupled system of integral equations by using the representations given in \cref{eq:rep_f_i}.
    By discretizing the function it turns into an eigenvalue problem which is solved numerically.
    Based on the results for the Faddeev amplitudes the wave function can be calculated, details
    can be found in \cref{ap:wf_heft}.
    We check the convergence of the results for the wave functions and other quantities by varying the number
    of mesh points used for this discretization and of the mesh points used for subsequent integrations.
    
    In addition to the scale $\beta$ that parameterizes the scale at which the regulator function \(\reg{{}}{p}\) cuts off the two-body
    t-matrix in Eq.~(\ref{eq:t_2bd}) we also place a cutoff $\Lambda$ on the 
    momentum-space integral equations obtained from \cref{eq:f_i} by using \cref{eq:rep_f_i}.
    We vary these two-body and three-body cutoffs and assess how sensitive our predictions are to that variation.
    Typically we use the same value for both cutoffs.

\section{\texorpdfstring{Ground-state \(\boldsymbol{nn}\) relative-momentum distribution}{Ground-state nn relative-momentum distribution}}\label{sec:nn_gs_distrib}

We compare ground-state \(nn\) relative-momentum distributions obtained with Halo EFT and with \lgmacronym.
By doing so, we can analyze and understand the uncertainty in the ground-state momentum distribution, which is an
important ingredient for the final distribution after the knockout.
As a preparation, we discuss the details of our definition of the distribution.

In the beginning of \cref{sec:tb_calcs}, the reference states for obtaining
wave functions in a partial-wave basis were discussed. Symmetry considerations yielded that only \(\Omega_c^{(0,l,l)}\) (\(l\) is even) and
\(\Omega_c^{(1,l,l)}\) (\(l\) is odd) are relevant. Using these different reference states yields complementary information
due to the orthogonality of their angular and spin part.
We calculated wave functions in our leading-order Halo EFT framework using both sets of states for low \(l\) in the low-energy region up to roughly \SI{140}{\mev}.
As expected, the importance of wave-function components decrease with increasing \(l\).
And in fact, only the wave-function component with partial-wave quantum numbers \(\Omega_c^{(0,0,0)}\) is relevant in this region.
All other components are suppressed in this region by a factor of approximately 20, or even more. 

Therefore in what follows we define the wave function
\begin{equation}
    \Psi_c{\K{p,q}} \coloneqq \ibraket{c}{p,q;\Omega_c^{(0,0,0)}}{\Psi}{} \,.
\end{equation}
For simplicity we sometimes use the abbreviated symbol \(\Omega_c \coloneqq \Omega_c^{(0,0,0)}\) in what follows. 
The corresponding ground-state \(nn\) relative-momentum distribution is given by
\begin{equation}
    \rho{\K{\pnn}} \coloneqq \rint{\q} \pnn^2 \left| \Psi_c{\K{\pnn,q}} \right|^2 \,.
\end{equation}

In the case of \lgmacronym, the suppression of the other wave functions compared to the wave function
with the quantum numbers \(\Omega_c^{(0,0,0)}\) is generally not as strong as in case of the LO Halo EFT calculation.
Nevertheless, in order to do an appropriate comparison, also for the \lgmacronym~we calculate only 
this wave-function component.
This choice is also motivated by the fact that after the \(nn\) final state interactions the dominance of the component \(\Omega_c^{(0,0,0)}\),
where the \(nn\) pair is in \(^1S_0\), is even increased in the low-energy region, as the \(nn\) interaction is much stronger in this partial wave.

Note that we have developed a cross-check for our results for ground-state \(nn\) relative-momentum distributions
\(\rho{\K{p_{nn}}}\) obtained in LGM. We calculate \(\expval{r_{nn}^2}\) for different partial waves \(l\) using the relation
\begin{equation}\label{eq:rsq_from_distrib}
  \expval{r_{nn}^2}_l = - \frac{\pi}{4} \K{ \int \dd{p_{nn}} \partial_{p_{nn}}^2 \rho_l{\K{p_{nn}}} - 2 \K{ 1 + l\K{l+1} } \int \dd{p_{nn}} \frac{\rho_l{\K{p_{nn}}}}{p_{nn}^2} }
\end{equation}
and compare the overall result with the results from Ref. \cite{Zhukov:1993aw},
where \(\expval{r_{nn}^2}\) and other observables were obtained in similar model calculations.

\subsection{Comparison of results}
\begin{figure}[htb]
    \centering
    \includegraphics[width=.47\textwidth]{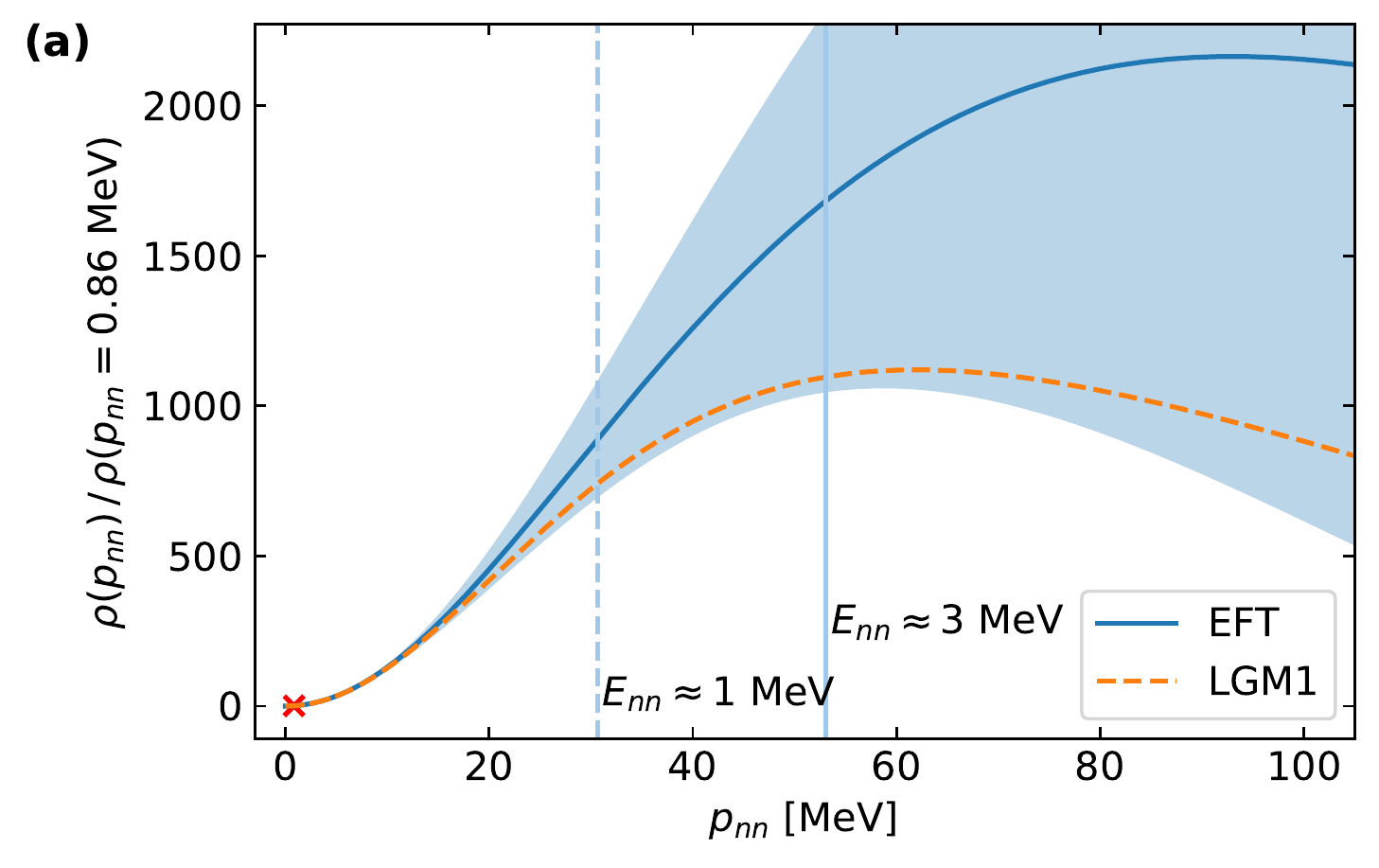}
    \includegraphics[width=.47\textwidth]{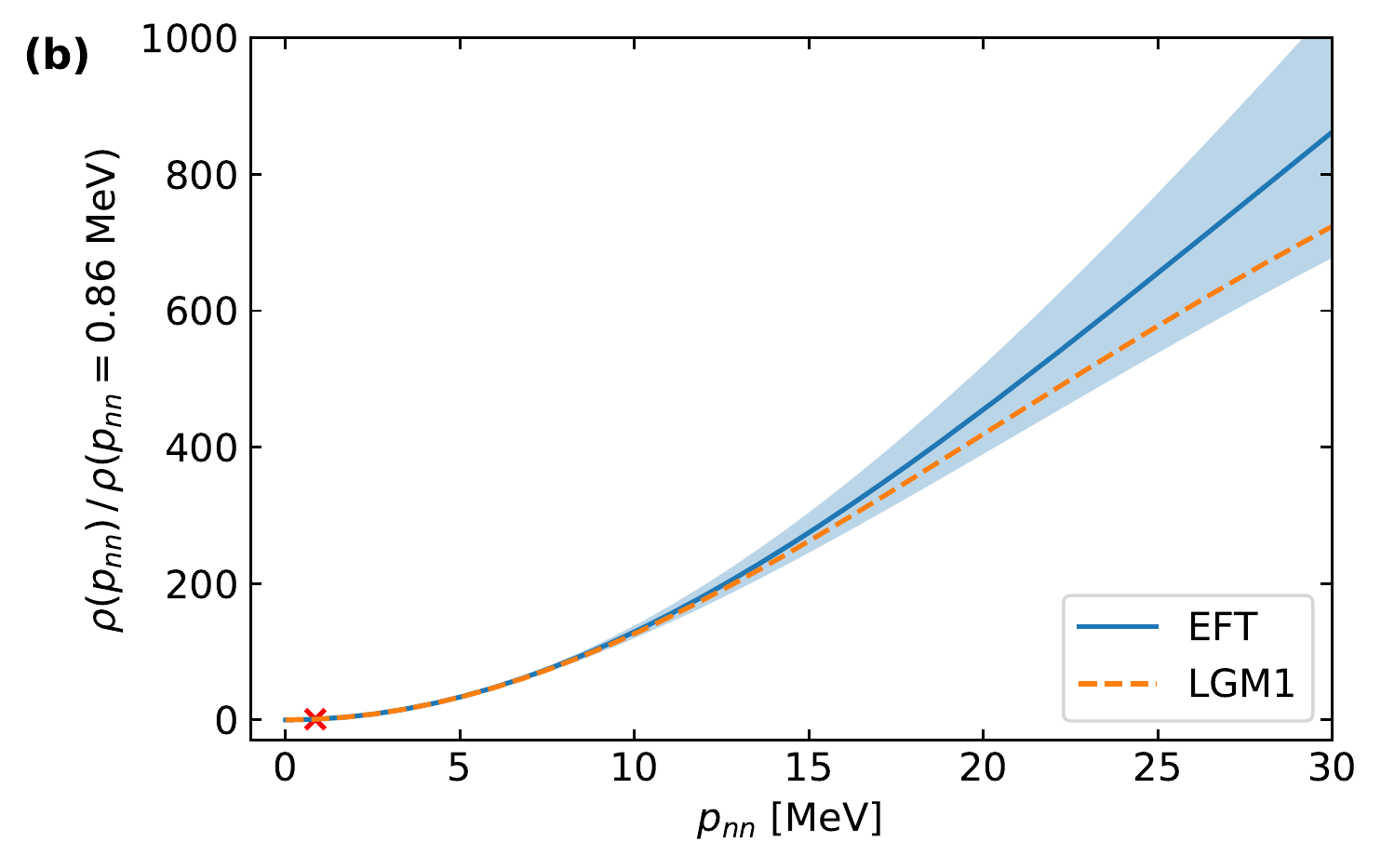}
    \caption{\lgmacronym~result in comparison with the Halo EFT result. 
    The three-body potential in the \lgmacronym~calculation was tuned to reproduce \(\btz\).
    The used settings for \lgmacronym~are denoted by \lgmacronym 1.
    In order to be independent of the normalization the distributions are divided by their value at a
    certain position, which is indicated by the red cross.
    Note the dashed and solid vertical lines in the left panel (a).
    They indicate relative energies of \SI{1}{\mev} and \SI{3}{\mev} respectively.
    In the planned experiment \SI{1}{\mev} will be roughly the upper bound of the measurement range.
    }\label{fig:boris1}
\end{figure}
\Cref{fig:boris1} shows the \lgmacronym~result for the ground-state \(nn\) relative-momentum distribution in
comparison with the leading-order Halo EFT result.
They are normalized to have a certain arbitrary value at a certain position.
We use this normalization procedure, as the absolute value is not necessary for determining the scattering length.
More information on that can be found in \cref{sec:nn_scat_length}.
This also avoids the difficulty that the norm of the EFT results depends on values of the wave function
outside its range of validity.
The uncertainty band of the EFT result is based on the size of the next-to-leading-order corrections.
They are suppressed by \(p/\mhi\).
Accordingly the uncertainty is \(\Delta \rho{\K{p}} \approx \rho{\K{p}} \frac{p}{\mhi}\).
In the left panel (a) it can be seen that EFT and \lgmacronym~results agree within the uncertainty bands of the LO Halo EFT result.
The EFT distribution has generally bigger values and its maximum is at higher momenta.
In higher-order Halo EFT calculations, the uncertainty bands (i.e. the relative uncertainty) will get smaller.
Agreement between that higher-order Halo EFT calculation and \lgmacronym~result is expected within this smaller uncertainty band.
That implies that next-to-leading-order (NLO) corrections will move the Halo EFT result towards the \lgmacronym~one, although at some high order 
no further improvement of agreement can be expected. Eventually, 
the assumptions of the model calculation will become visible in terms of small insurmountable
differences between a high-order EFT and a model calculation.
The right panel (b) shows that the agreement is, as expected, better in the low-energy region.
Importantly, the determination of the \(nn\) scattering length involves measuring  the distribution only up to $E_{nn} \approx 1$ MeV, 
i.e. the region where the agreement is especially good.

\begin{figure}[H]
  \centering
  \includegraphics[width=.47\textwidth]{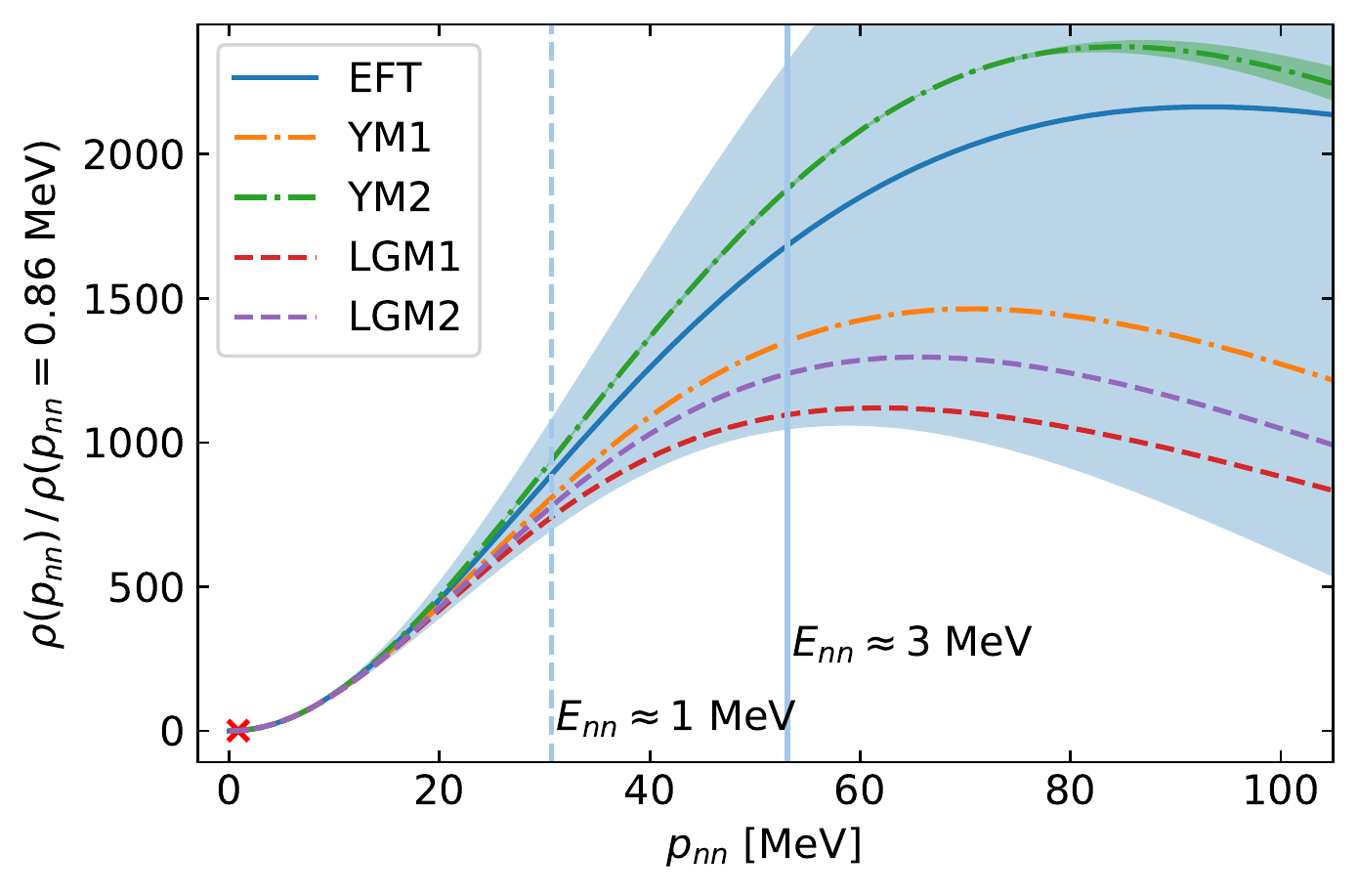}
  \caption{YM results (dot-dashed lines) in comparison with the LO Halo EFT result (solid line) as well as LGM results (dashed lines).
  All are normalized to have a certain arbitrary value at a
  momentum indicated by the red cross. Meanwhile, 
  the vertical lines dashed and solid indicate, respectively, relative energies of \SI{1}{\mev} and \SI{3}{\mev}.
  The estimated numerical uncertainties of the YM1 (YM2) result is indicated by the dark blue (green) band.
 (Not all are large enough to be visible.)
 The estimation is based on the comparison of the calculation with a three-body cutoff at momenta of
 \SI{2250}{\mev} with one with a cutoff of \SI{1500}{\mev} and half as many mesh points. 
The light blue error band for the 
  Halo EFT result shows the expected size of the NLO correction.
  }\label{fig:ym_eft_lgm}
\end{figure}

This comparison shows the consistency of the results. But we are interested in the 
sources of the discrepancies and in what results we can expect from
a Halo EFT calculation at next-to-leading  order.
There are several possible sources for the discrepancies, such as: the phenomenological \lgmacronym~three-body potential,
different effective-range-expansion (ERE) parameters, or different off-shell properties of the two-body interactions
that are not compensated by the used three-body forces (see Ref. \cite{Polyzou:1990} for details on this topic).
Additionally, the discrepancies can be caused by terms which are part of higher-order EFT descriptions, e.g.,
the unitarity term of the \(nc\) system or interactions in additional partial waves such as \(^2P_{1/2}\) and \(^2S_{1/2}\) in the \(nc\) system.
In order to estimate the importance of these different effects, we performed additional model calculations.
We introduced modified versions of \lgmacronym~that have fewer \(nc\) interaction channels (\lgmacronym2) or in which the three-body potential is  completely absent or of shorter range.
The \lgmacronym2 calculations show that the \lgmacronym~result gets more similar to the EFT one if the \(nc\) interactions are turned off in channels other than the \(^2P_{3/2}\) (cf. Fig.~\ref{fig:ym_eft_lgm}).
The \(s\)-wave, \(d\)-wave and \(^2P_{1/2}\) \(nc\) interactions, which in the EFT are higher-order effects, are therefore causing part of the discrepancy.
Meanwhile, calculations using other \lgmacronym~variants (see supplemental material \cite{supp_mat}) show that the phenomenological \lgmacronym~three-body force is an important ingredient:
if it is omitted, the gap with the EFT result increases.
However, the range of the \lgmacronym~three-body force seems to play only a small role. Using a value of 2.5 fm instead of 5 fm for \(\rho_\mathrm{3B}\) has only a small effect on the 
\lgmacronym~result, provided the three-body force's strength 
is adjusted to reproduce the physical binding energy of the three-body system.

While the interactions for the \lgmacronym~calculations are specified in terms of coordinate-space matrix elements of potentials,
momentum-space t-matrices are used for the EFT calculations.
In order to connect the EFT calculations to a model, we also performed calculations using a model formulated
directly in momentum space.
This can be achieved by using our computer code for the EFT calculations with different separable t-matrices.
The resulting model calculation has the same interaction channels as Halo EFT at LO and is similar to those of Refs. \cite{Hebach:1967bpg,Shah:1970wu,Ghovanlou:1974zza}.
We chose separable t-matrices with Yamaguchi form factors, with interaction parameters adjusted to reproduce effective-range-expansion
parameters. This yields reasonable phase shifts.
In a first step, we compared the Yamaguchi model (YM) results with our EFT and \lgmacronym~results.
We found that the YM results for the ground-state momentum distribution are much more similar to the results from \lgmacronym2 (\lgmacronym~with the 
reduced set of interaction channels) than to the EFT results.
This implies that if we understand the discrepancy between YM and EFT we also understand the discrepancy between \lgmacronym~and EFT.

\cref{fig:ym_eft_lgm}, shows the standard \lgmacronym~and YM calculations, labeled
\lgmacronym1 and YM1. The \lgmacronym~calculation with a reduced set of channels (\lgmacronym2) is also shown. 
In addition, we perform a YM calculation with the unitarity term of the \(nc\) t-matrix removed, while other higher-order terms which are part of the YM but are not in the
LO Halo EFT calculation are retained (YM2). The YM2 calculation comes out quite close to the LO Halo EFT result, indicating that the unitarity term in the \(nc\) t-matrix, which 
is an NLO effect in the EFT, has a significant influence on the ground-state momentum distribution and causes a large fraction of the YM-EFT difference.
This implies that the NLO Halo EFT calculation will likely agree much better with a YM or \lgmacronym~calculation than
the LO Halo EFT does.

To conclude this subsection: the comparison between EFT and \lgmacronym1 yields agreement at the expected level: the EFT uncertainty bands are indeed robust.
We expect the NLO Halo EFT result to be closer to this model that includes additional effects, and we tracked down the specific NLO term that should most improve agreement.
Comparisons with additional model calculations indicate that 
the unitarity
term of the \(nc\) interaction plays a significant role in this distribution. 
A more detailed analysis of the differences and additional plots
can be found in the supplemental material \cite{supp_mat}.

\subsection{\texorpdfstring{Influence of the \(\boldsymbol{nn}\) scattering length on the ground-state momentum distribution}{Influence of the nn scattering length on the ground-state momentum distribution}}
\begin{figure}[H]
    \centering
    \includegraphics[width=.5\textwidth]{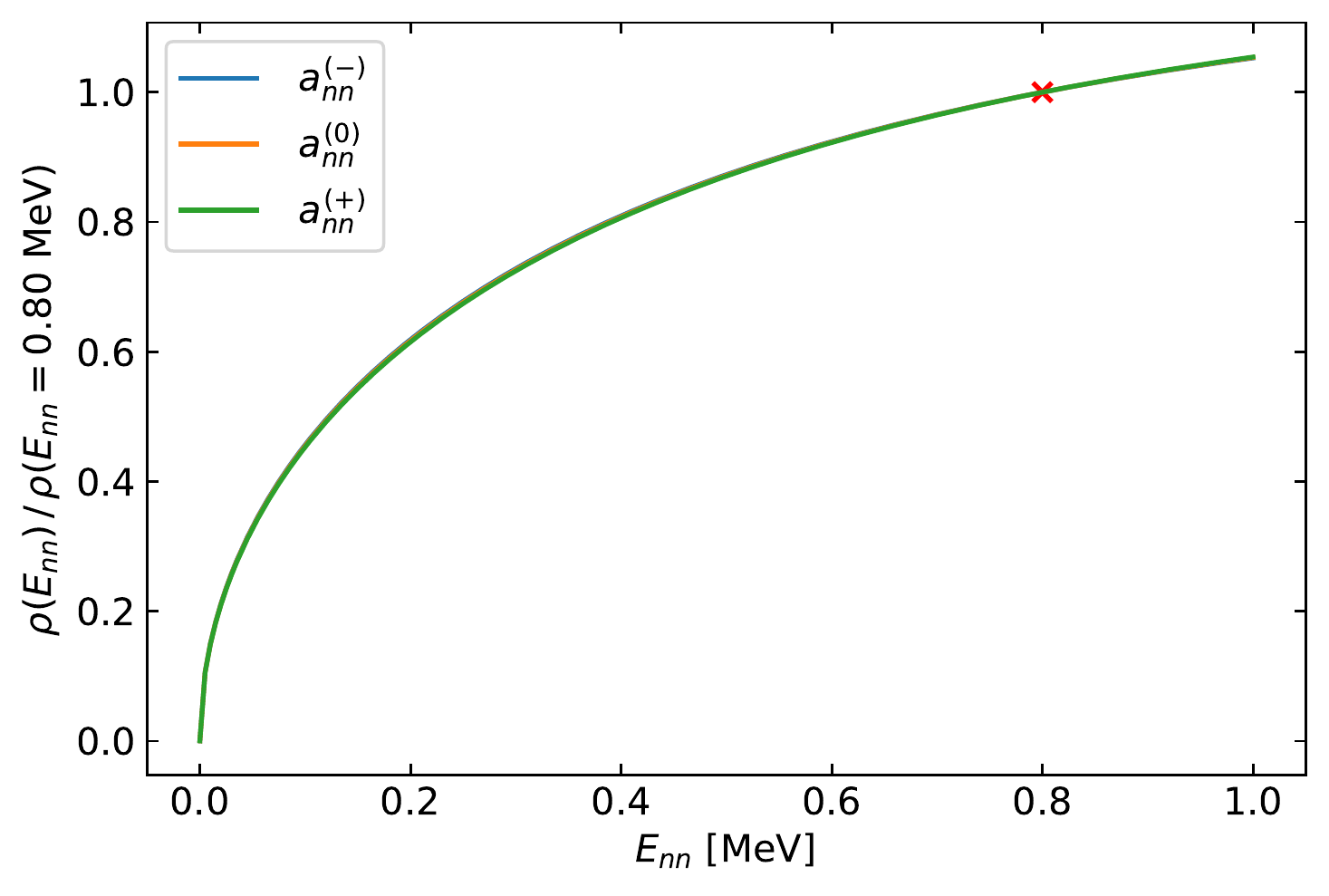}
    \caption{Ground-state \(nn\) relative-energy distributions for different \(nn\) scattering lengths.
    The definitions \(a_{nn}^{(+)} = \SI{-16.7}{\fm}\), 
    \(a_{nn}^{(0)} = \SI{-18.7}{\fm}\) and \(a_{nn}^{(-)} = \SI{-20.7}{\fm}\) hold.
    All results are based on \(\Psi_c{\K{p,q}}\).
    \lbh~was used. Based on a comparison with a calculation with
    half as many mesh points and \lbl~the numerical uncertainty is negligible.
    In order to be independent of the normalization the distribution is divided by its value
    at a certain position, which is indicated by a red cross.}\label{fig:gs_en_distribs}
\end{figure}

Up to this point, we have compared different ground-state \(nn\) relative-momentum distributions.
In the next section, we will show \(nn\) relative-energy distributions after taking \(nn\) final state interactions
into account.
Before we do that, we want to show what an intermediate step of this procedure looks like\footnote{
    Note, that strictly speaking the ground-state relative momentum or energy distribution is only an intermediate step
    in an enhancement factor based FSI approach.
    If the t-matrix itself is used, it has to be applied at the wave-function level.
    Before and after its usage the respective distributions can be calculated, but in this approach one
    cannot get directly from the ground-state momentum or energy distribution to the one after \(nn\) FSI.
    The details can be found in the next section.
}.
We show the ground-state \(nn\) relative-energy distribution \(\rho{\K{E_{nn}}}\) with
\(E_{nn} = p_{nn}^2 /\K{2\mu_{nn}}\).
Especially, we want to investigate the influence of \(a_{nn}\) on this distribution.
The relation between the momentum and the energy distribution is
\begin{equation}\label{eq:mom_distrib2en_distrib}
    \rho\K{E_{nn}} = \sqrt{\frac{\mu_{nn}}{2E_{nn}}} \rho\K{\sqrt{2\mu_{nn}E_{nn}}}\,,
\end{equation}
where we use the common style to distinguish the different variants of the function, i.e., the different functions differ only by their arguments.
The necessity of the additional factor can be seen from dimensional analysis.
The factor follows from a substitution in the normalization integral of the distribution.
The normalization condition reads
\( \int \dd{E_{nn}} \rho\K{E_{nn}} = 1\,.\)
We plot the distribution obtained with different \(nn\) scattering lengths in \cref{fig:gs_en_distribs}.

It can be seen that the shape of the relative-energy distribution is different from the one of the momentum-
distribution.
Additionally, we observe that the influence of the \(nn\) scattering length on the ground-state distribution is
negligible.
In contrast, we will see that, after taking the \(nn\) FSI into account, the distribution is sensitive to the scattering length.
Thus, \hesix~serves in the proposed experiment as a source of low-energy neutrons.
Its structure is not sensitive to \(a_{nn}\).
It is the final state \(nn\) interaction that enables the measurement of the scattering length.

Quantitative information on the negligible influence of the scattering length on the ground-state distribution
can be found in the supplemental material \cite{supp_mat}.
This plot shows the ratios of the distributions obtained with different scattering lengths.
They agree to better than 1\%.

\section{\texorpdfstring{\(\boldsymbol{nn}\) relative-energy distribution after knockout}{nn relative-energy distribution after knockout}}\label{sec:final_nn_distrib}
The next step is to calculate the \(nn\) relative-energy distribution after the knockout reaction, in which the \(\alpha\) core
of \hesix~is removed via a collision with a proton. In the experiment under discussion in this paper the knockout takes place
in inverse kinematics, with a beam of \hesix~nuclei impinging on a hydrogen target~\cite{nn_scat_len_ribf_prop2018}. 
We employ the sudden approximation, i.e., we assume that the reaction mechanism is rapid, quasi-free knockout of the \(\alpha\) and subsequent interactions between
neutrons and the \(\alpha\) or the proton that struck the \(\alpha\) can be neglected. 
Therefore, in our analysis it is sufficient to treat the potential causing the knockout as an external potential.
The Hilbert space for our problem is then a three-body \(\alpha n n\) Hilbert space. Note that as an alternative to this effective three-body treatment a four-body description of this reaction is possible. 
The proton, which causes the knockout, would then be explicitly included in the Hilbert space.
But, in our three-body treatment, that proton merely generates a potential that enables the production of the final state. We therefore refer to this as the production potential \(V\).
The quantitative properties of the final state are influenced by the \(nn\) potential, so that we face
a two-potential scattering problem. 
A comprehensive discussion of such problems can be found in Refs. \cite{goldberger2004collision,GellMann:1953zz}.

Before going into the details of that two-potential formalism and its application to \(\alpha\)-particle knockout in \hesix, we
want to discuss some fundamental aspects of the \(^6{\rm He}(p,p\alpha)\) reaction.
The initial state is the ground state of \hesix~denoted by \(\ket{\Psi}\). It fulfills 
the Schr\"odinger equation
\begin{equation}\label{eq:sdeq_initial_state}
  \K{ K_{nn} + K_{(nn)c} + V_{nn} + V_{nc} + V_{3B} } \ket{\Psi} = -B_3^{(0)} \ket{\Psi} \,,
\end{equation}
where the kinetic energy operators are denoted by \(K\) and \(V_{nc}\) represents the interaction of the core
with each of the two neutrons.
In this experiment the final state is measured by a  setup which detects a free \(nn\) state with definite relative momentum. Meanwhile, the 
 \(\alpha\) particle is detected at a very different angle where it is not interacting with the \(nn\) pair. Because of the high initial velocity of \hesix~the neutrons will leave the \(\alpha\) particle and proton
after their interaction quickly behind, as in the lab frame the \(nn\) pair travels at almost the initial velocity of the \hesix~beam.

Accordingly, we consider final states parameterized by the momenta \(p\) and \(q\), that fulfil the free Schr\"odinger equation
\begin{equation}
  \K{K_{nn} + K_{(nn)c}} \iket{p,q}{c} = \K{-\btz + E_\mathrm{KO}} \iket{p,q}{c} \,,
\end{equation}
where the energy transfer from the knockout \(E_\mathrm{KO}\) that is delivered by the proton:
\begin{equation}
  -\btz + E_\mathrm{KO} = \frac{p^2}{2 \mu_{nn}} + \frac{q^2}{2 \mu_{(nn)c}} \,.
\end{equation}
In order for the final state to be a scattering state, the condition \(E_\mathrm{KO} > \btz\) must be fulfilled.
Note that energy is 
still conserved in the four-body (\(p\alpha nn\)) system. But the energy
 \(E_\mathrm{KO}\) describes a transfer of energy into the internal (not center-of-mass) energy of the three-body
  system.  \(E_\mathrm{KO}  < E_{\mathrm{lab},{}^6\mathrm{He}}\) then holds, where \(E_{\mathrm{lab},{}^6\mathrm{He}}\)
  is the initial kinetic energy of the \hesix~projectile in the laboratory frame. 

Since we assume the proton interacts only with the \(\alpha\) particle we adopt a 
production potential \(V\) that does not change the relative momentum
of the \(nn\) pair:
\begin{align}\label{eq:decoupling_prod_pot}
  \imel{c}{p,q}{V}{\Psi}{}
  &= \rint{\pp} \rint{\qp} \braket{p}{p'} \mel{q}{\tilde{V}}{q'} \ibraket{c}{p',q'}{\Psi}{}
  = \rint{\qp} \mel{q}{\tilde{V}}{q'} \ibraket{c}{p,q'}{\Psi}{} \,.
\end{align}
In other words, we assume a factorization of the production potential into a \(nn\) part and a \((nn)c\) part with
the \(nn\) part being an identity operator: \(V = \id \otimes \tilde{V}\)\,.

We now make use of the formalism of Ref. \cite{goldberger2004collision} for scattering from two potentials.
A more detailed summary of this formalism can be found in the supplemental material \cite{supp_mat}.
The two potentials are taken to be the production potential \(V\) and 
the potential \(U\) causing the final-state interactions.
We make use of a helpful identity of two-potential scattering theory
for calculating the probability amplitude of the transition from a state \(\ket{\alpha}\) to a state \(\ket{\beta}\):
\begin{equation}
  T_{\beta \alpha} = \mel{\beta}{T_{U+V}^{(+)}}{\alpha} \,.
\end{equation}
These states satisfy the equations \(H_0 \ket{\alpha} = E_\alpha \ket{\alpha}\) and \(H_0 \ket{\beta} = E_\beta \ket{\beta}\) 
with \(E_\alpha = E_\beta = E\). 
\(H_0\) should be thought of as the part of the Hamiltonian that does not
include the interactions \(U\) and \(V\).
The operator \(T_{U+V}^{(+)}\) is then the t-matrix for scattering involving \(U\) and \(V\). 
It satisfies the standard Lippmann-Schwinger equation, where the potential is given by \(U+V\).
It is possible to
dissect this transition amplitude (as well as this overall t-matrix itself) into two terms
by using M{\o}ller operators~\cite{goldberger2004collision} .
One of the two terms contributes for elastic scattering reactions because \(V\), 
which causes the production of the final state, is missing there.
We are not interested in elastic scattering here and so focus on the other term. We consider the situation in which the production potential produces a transition from a bound state to a scattering state in a subsystem.
In this case \(V\)
induces a transition to an eigenstate of \(H_0\) that is orthogonal to the initial eigenstate \(|\alpha \rangle\).
While this discussed relation is generally a suitable starting point for the calculations,
for our application we have to modify it to accommodate the case that 
the final-state interaction \(U\) is
part of the Hamiltonian that describes the initial state.
In such a case the stationary Schr\"odinger equation for the initial state becomes \(\K{H_0 + U} \ket{\alpha} = E_\alpha \ket{\alpha}\).
We continue to assume a free final state, so \(H_0 \ket{\beta} = E_\beta \ket{\beta}\) stays unchanged.
Goldberger and Watson show in Ref. \cite{goldberger2004collision} that under these assumptions we have 
\begin{equation}\label{eq:Tba_only_first_term_sc}
  T_{\beta \alpha} = \mel{\beta}{ \K{\Omega_U^{(-)}}^\dagger V \K{  \id + \K{E-K-U-V + \ci \epsilon}^{-1} V } }{\alpha} \,,
\end{equation}
where the M{\o}ller operator corresponding to the potential \(U\) is denoted by \(\Omega_U^{(-)}\).

Now we have to evaluate \cref{eq:Tba_only_first_term_sc}.
While we can (and will) evaluate it directly using the already mentioned assumption about \(V\), this expression has
also often been evaluated via final-state interaction (FSI) enhancement factors.
In the next subsection we give a brief overview of this approach.

\subsection{FSI enhancement factors}
    The FSI enhancement factors are a technique for approximately calculating the effect of the final-state interaction
    on the transition probability. The production potential is not explicitly taken into account.
    These enhancement factors as a generic tool were introduced
    by Watson \cite{Watson:1952ji} as well as by Migdal \cite{Migdal:1955_1}.
    Watson used the approach of two-potential scattering theory to derive a relation similar to \cref{eq:Tba_only_first_term_sc} and from it the enhancement factor.
    A detailed explanation of this way of establishing enhancement factors can be found in Ref. \cite{goldberger2004collision}.

    In this context, it is important to note that enhancement factors were introduced
    for describing reactions such as \( \pi^- + d \to n + n + \gamma\) \cite{Watson:1951}.
    Here the \(nn\) enhancement factor enters in a fundamentally different way than it does in \hesix. In
    the radiative pion capture reaction 
    the production potential and the final-state interaction both affect the same subsystem, i.e., the \(nn\) system in this case.
    In contrast when a high-momentum \hesix~impinges on a proton target the production potential acts in a different subsystem than 
    does the final-state interaction. Here we first discuss the original use case, where both \(V\) and \(U\) act on the same
    subsystem. 
    We then discuss the implications for how these factors should be computed in the case of the reaction we are interested in.
        
    The enhancement factors can be derived from \cref{eq:Tba_only_first_term_sc} by using a state of definite momentum
    as the final state and the bound state as the initial state: \(\ket{\beta} = \ket{\v{p}}\) and \(\ket{\alpha} = \ket{\Psi}\).
    During this derivation it is assumed that the production potential is weak and so the operator given by the expression in the brackets to the 
    right of the first \(V\) in \cref{eq:Tba_only_first_term_sc} can be approximated by \(\id\).
    Additionally, it is assumed that production potential is local and only \(s\)-wave interactions are taken into account.
    Furthermore, it is required that the initial-state wave function and/or the production potential peak at short distances.
    If these conditions are satisfied one arrives at the following expression for the final momentum distribution:
    \begin{equation}\label{eq:rho_and_ef}
      \rho^{(G_i)}{\K{p}} \propto G_i{\K{p}} \rho{\K{p}}\,,
    \end{equation}
    where \(G_i{\K{p}}\) is the enhancement factor and
    \(\rho{\K{p}}\) is the momentum-space probability distribution from the initial (bound) state.
    Note that to obtain this expression we assumed that the production potential does not alter the momentum \(p\). 
    For the application we have in mind here this assumption holds, because the production potential and FSI potential act in different subsystems.
    
     Different enhancement factors can be derived depending on the particular assumptions made, especially in regard to the short-distance behavior
    of the production potential and/or initial-state wave function.
    This is why we added the index \(i\) to the enhancement factor \(G_i{\K{p}}\).
    A common variant of this enhancement factor, derived in Ref. \cite{goldberger2004collision},
    is \footnote{
      Note, that in Ref. \cite{goldberger2004collision} the enhancement factor has \(1/a_{nn}\) instead of \(-1/a_{nn}\)
      in the denominator. This is rooted in a different sign convention for the scattering length.
      We define \(k\cot{\K{\delta_0{(k)}}} = -1/a_0 + r_0 k^2  /2 + \mathcal{O}{\K{k^4}}\).
    }
    \begin{equation}\label{eq:G1}
      G_1{\K{p}} = \frac{ \K{\K{p^2 + \alpha^2} r_{nn}/2 }^2 }{ \K{-\frac{1}{a_{nn}} + \frac{r_{nn}}{2}p^2 }^2 + p^2 } \,,
    \end{equation}
    where \(\alpha=1/r_{nn} \K{1 + \sqrt{1 - 2 r_{nn}/a_{nn}} } \).
    This enhancement factor is based on the assumption that \(V\ket{\Psi}\) peaks at \(r=0\).
    It is also possible to derive enhancement factors for the case
    that \(V\ket{\Psi}\) peaks at some other radius \(\tilde{r}\).
   Further discussion regarding the derivation of the enhancement factor and how to obtain it for a
    general \(\tilde{r}\) can be found in the supplemental material \cite{supp_mat}.
    
    So far this discussion of enhancement factors focused on two-body systems.
     To close this section we point out that this formalism can also be used in \(n\)-body systems.
    That extension assumes that the FSI is a two-body interaction within one specific particle pair; the requirement regarding the
    short-distance behavior then applies to the corresponding two-body subsystem of the \(n\)-body state.
    For a system with  \(n > 2\) the  \(\rho{\K{p}}\) in \cref{eq:rho_and_ef} is the momentum-space
    probability distribution of the bound state after all other momenta are integrated out. Furthermore, since the FSI enhancement factor factorizes the FSI from the action of the production potential, it can be used not only
    in the case where the production potential acts in the same subsystem as the FSI potential,
    but also in cases where the two act on different subsystems of the overall \(n\)-body system.
    
  \subsection{Explicit calculation of rescattering}
    Having discussed the FSI enhancement factors in the previous subsection, we now turn our attention to the direct calculation of the wave function 
    after FSI.
    Our starting point is again \cref{eq:Tba_only_first_term_sc} except that now we consider it in the context of the breakup of a three-particle state into an \(nn\) pair 
    and a residual cluster, like an \(\alpha\) particle. For concreteness we consider
    the final state \(\bra{\beta}\) to be the free state of the \(nn\) pair and the \(\alpha\) particle and specify that state via the relative momentum within the \(nn\) pair, \(p\),
    and the momentum of the \(\alpha\) particle relative to the \(nn\) pair, \(q\), as well as the partial-wave quantum numbers
     \(\Omega\). The state is \(\ibra{c}{p,q; \Omega }\).
    The initial state \(\ket{\alpha}\) is given by the \hesix~bound state \(\ket{\Psi}\).
    Using the notation of \cref{eq:sdeq_initial_state}, this implies that the final state is an eigenstate
    of \(H_0 = K_{nn} + K_{(nn)c}\), while the initial state is an eigenstate of
    \(H_0 + V_{nn} + V_{nc} + V_{3B}\).
    That implies that the FSI potential \(U\) is given by \(V_{nn} + V_{nc} + V_{3B}\). 
    This reflects the fact that in addition to \(nn\) interactions also \(nc\) interactions as well
    as three-body interactions are possible final-state interactions happening after the knockout.
    However, due to the kinematics of the reaction and the halo structure of \hesix,~final-state
    \(nc\) or three-body interactions should be strongly suppressed.
    Accordingly, in the context of this calculation we approximate \(\Omega_U^{(-)}\) by \(\Omega_{V_{nn}}^{(-)}\).
    We obtain
    \begin{equation}\label{eq:T_He6}
      \mathcal{T}_\Omega{\K{p,q}} = \imel{c}{ p,q; \Omega }{ \K{\Omega_{U}^{(-)}}^\dagger V \K{\id + \K{E-K-U-V + \ci \epsilon }^{-1} V} }{\Psi} \,.
    \end{equation}
    where \(U\) is to be approximated by the \(nn\) potential. 
    For the energy \(E\) of the M{\o}ller operator we have to insert the energy of the final state 
    \(p^2 / \K{2\mu_{nn}} + q^2 / \K{2\mu_{(nn)c}}\).
    Since the FSI potential \(U\) is approximated by \(V_{nn}\) and thereby acts only in the \(nn\) subsystem, we can make use of the identity
    \begin{align}
      \Omega_U^{(\pm)} \iket{p,q; \Omega }{c} &= \Ke{ \id + \K{ p^2 / \K{2\mu_{nn}} + q^2 / \K{2\mu_{(nn)c}} - K_{nn} - K_{(nn)c} - U \pm \ci \epsilon }^{-1} U } \iket{p,q; \Omega }{c} \nonumber \\
      &= \Ke{ \id + \K{ p^2 / \K{2\mu_{nn}} - K_{nn} - U \pm \ci \epsilon }^{-1} U } \iket{p,q; \Omega }{c} \,, 
    \end{align}
    i.e., we use the fact that \(V_{nn}\) commutes with \(K_{(nn)c}\), and so \(K_{(nn)c}\) can be replaced by its eigenvalue for the state
    \(\iket{p,q;\Omega}{c}\). 
    Next, since the production potential is assumed to be weak, in \cref{eq:T_He6} we use only the lowest order
    of the operators next to \(\K{\Omega_U^{(-)}}^\dagger\), i.e., retain only the identity operator in the round brackets to the right of \(V\) in Eq.~(\ref{eq:T_He6}). 
    Furthermore, we assume that \(V\) decouples as formulated in \cref{eq:decoupling_prod_pot}.
    It is then useful to express the M{\o}ller operator in terms of the t-matrix according to\footnote{
      We use here that \(\Omega_U^{(-)}\) acts on an eigenstate of \(H_0\).
    }
    \begin{equation}
      \K{\Omega_U^{(-)}}^\dagger = \id + \K{G_0^{(-)} t_U^{(-)}}^\dagger \,.
    \end{equation}
    We set \(\tilde{V}\) to \(\id\), which implies that the momentum \(q\) in \(\mathcal{T}_\Omega{\K{p,q}}\) 
    is the \(\alpha(nn)\) relative momentum before the reaction. We therefore
    calculate the probability amplitude as a function of the \(nn\) relative momentum after the reaction and the 
    \(\alpha(nn)\) relative momentum before the reaction.
    Another implication of not using an explicit expression for \(\tilde{V}\)
    is that we don't take into account that the overall probability of the knockout is smaller than 1.
    The implications of this on the analysis are discussed in \cref{ssec:cmp_results}, and will be accounted for
    by not trying to compute the absolute number of \(nn\) pairs produced, but only the shape of the distribution. 
    
    Under these assumptions the probability amplitude \(\mathcal{T}_\Omega{\K{p,q}}\), which can also be seen as a final-state wave function
    in an arbitrary partial wave \(\Omega\) after knockout and FSI, is given by
    \begin{align}\label{eq:psi_fsi}
      \Psi_c^{\K{\mathrm{wFSI};\Omega}}{\K{p,q}} &= \imel{c}{p,q;\Omega}{ \K{ \id + t_{nn,\K{\Omega}_{nn}}{\K{E_p}} G_{0}^{(nn)}{\K{E_p}} } }{\Psi}{} \nonumber \\
      &= \rint{\pp} \imel{c}{p,q;\Omega}{ \K{ \id + t_{nn,\K{\Omega}_{nn}}{\K{E_p}} G_{0}^{(nn)}{\K{E_p}} } }{p^\prime,q;\Omega}{c} \ibraket{c}{p^\prime,q;\Omega}{\Psi}{} \,,
    \end{align}
    where the multi-index \(\K{\Omega}_{nn}\) is the \(nn\) part of the multi-index \(\Omega\).
    
    The \(nn\) FSI is only significant in the ${}^1$S$_0$ partial wave, so 
    we use only the wave function
    component \(\Psi_c{\K{p,q}} \coloneqq \ibraket{c}{p,q;\Omega_c}{\Psi}{}\) for calculating the wave function after FSI.
    The \(nn\) part of this wave function's multi-index is \(l=0, s=0\)
    corresponding to the \(^1S_0\) channel.
    Accordingly, to obtain results for \(\Psi_c^{\K{\mathrm{wFSI}}}{\K{p,q}}\) a version of \cref{eq:psi_fsi} specific to FSI in this \(nn\) partial wave is used:
    \begin{align}
      \Psi_c^{\K{\mathrm{wFSI}}}{\K{p,q}} 
      &= \Psi_c{\K{p,q}} + \frac{2}{\pi} g_0{\K{p}} \frac{1}{a_{nn}^{-1} - \frac{r_{nn}}{2} p^2 + \ci p} \rint{\pp} g_0{\K{\pp}} \K{ p^2 - \pp[2]  + \ci \epsilon }^{-1}  \Psi_c{\K{\pp,q}} \label{eq:tmb_fsi_intermed_step}\\
      &= \Psi_c{\K{p,q}} + \frac{2}{\pi} g_0{\K{p}} \frac{1}{a_{nn}^{-1} - \frac{r_{nn}}{2} p^2 + \ci p} \nonumber \\
      & \quad \times \Ke{ \int_0^\Lambda \dd{\pp} \frac{\pp[2] \Psi_c{\K{\pp,q}} - p^2 \Psi_c{\K{p,q}}}{ p^2 - \pp[2] } - 
      \K{ \frac{\ci \pi}{2} - \frac{1}{2} \ln{\K{\frac{\Lambda + p}{\Lambda - p}}} } g_0{\K{p}} p \Psi_c{\K{p,q}} }. \label{eq:tmb_fsi_concrete}
    \end{align}
   Note that \(\Psi_c{\K{p,q}}\) is
   the wave function corresponding to the momentum distribution computed in the previous section. 
    The last equality holds in case of Heaviside functions as regulators using the cutoff \(\Lambda\): \(g_l{\K{p}} = p^l \Theta{\K{\Lambda-p}}\).
    An auxiliary calculation can be found in the supplemental material \cite{supp_mat}.
    Note that in the calculation leading to Eq.~\eqref{eq:tmb_fsi_concrete} we included the effective-range term in the $nn$ t-matrix in order to check its influence.
 The $nn$ relative-energy distribution below $E_{nn}=1.0$ MeV that is obtained with the choice \(r_{nn}=0\) in the FSI $nn$ t-matrix differs only slightly from the distribution obtained when the nominal effective range of \(r_{nn}=2.73~{\rm fm}\) is used there. (Cf. Fig.~\ref{fig:fsi_rnn} in Appendix \ref{ap:add_plots}.)

    This procedure for calculating the FSI is common and inter alia used for pion capture reactions with deuterium, see, e.g., Ref. \cite{Golak:2018jje}.
    It is also similar to the coherent FSI three-body model for the sudden breakup of two-neutron halos in collisions
    with heavy targets developed in Ref. \cite{Yamashita:2005yh}.
    This model was compared to experimental data on the \(nn\) correlation in the breakup of $^{11}$Li \cite{mar-plb00,petra-np04}, $^{14}$Be \cite{mar-plb00,mar-prc01} and $^6$He \cite{mar-plb00}.
    Within our discussion we were able to show the close connection of this method to two-potential scattering theory.
    Additionally, we reviewed the specific approximations which were made.

    It is interesting to note that certain FSI enhancement factors can be derived 
    from \cref{eq:tmb_fsi_intermed_step} by approximating the integral that appears there. This is not really surprising, since enhancement factors and this more exact calculation are both based on the findings
    of two-potential scattering theory.
    Nevertheless, the derivation elucidates the relationship of the enhancement-factor and explicit-calculation approaches to \(nn\) FSI and provides a different
    perspective on the enhancement factors.
    It is discussed in the supplemental material \cite{supp_mat}.

    After applying the FSI, the absolute value of the wave function can be calculated, the integral measure applied 
    and the \(q\)-momentum can be integrated out
    in order to obtain the probability density distribution as a function of the \(nn\) relative momentum, \(p\).
    Note that taking \(\tilde{V}\) into account in \cref{eq:decoupling_prod_pot}, could distort the probability distribution in \(p\), even though \(\tilde{V}\)
    acts only in the \(q\) subspace of the Hilbert space.
    However, such effects are expected to be small, due to the kinematics of the proposed knockout reaction.
    The formula for the probability density \(\rho^{(t)}{\K{p}}\) in this approach (where the superscript denotes that the FSI is computed via the t-matrix) reads
    \begin{equation}
      \rho^{(t)}{\K{p}} = \intdd{\q} p^2 q^2 \left| \Psi_c^{\K{\mathrm{wFSI}}}{\K{p,q}} \right|^2 \,.
    \end{equation}
    The relative energy distribution can then be calculated from the momentum distribution by using \cref{eq:mom_distrib2en_distrib}.
    
    The density \(\rho^{(t)}{\K{p}}\) obeys the normalization condition \(\intdd{\p} \rho^{(t)}{\K{p}} = 1\).
    However, we remind the reader that the existence of other channels than this one is not taken into account in our calculation.
    Thus, this normalization condition does not represent the actual probability of knockout, which in reality will be \(< 1\).
    Additionally, this normalization condition requires that the wave function component in use, i.e., 
    \(\Psi_c{\K{p,q}} \coloneqq \ibraket{c}{p,q;\Omega_c}{\Psi}{}\), be normalized to one, which is another approximation.

  \subsection{Comparison of results}\label{ssec:cmp_results}
\begin{figure}[H]
      \centering
      \includegraphics[width=.5\textwidth]{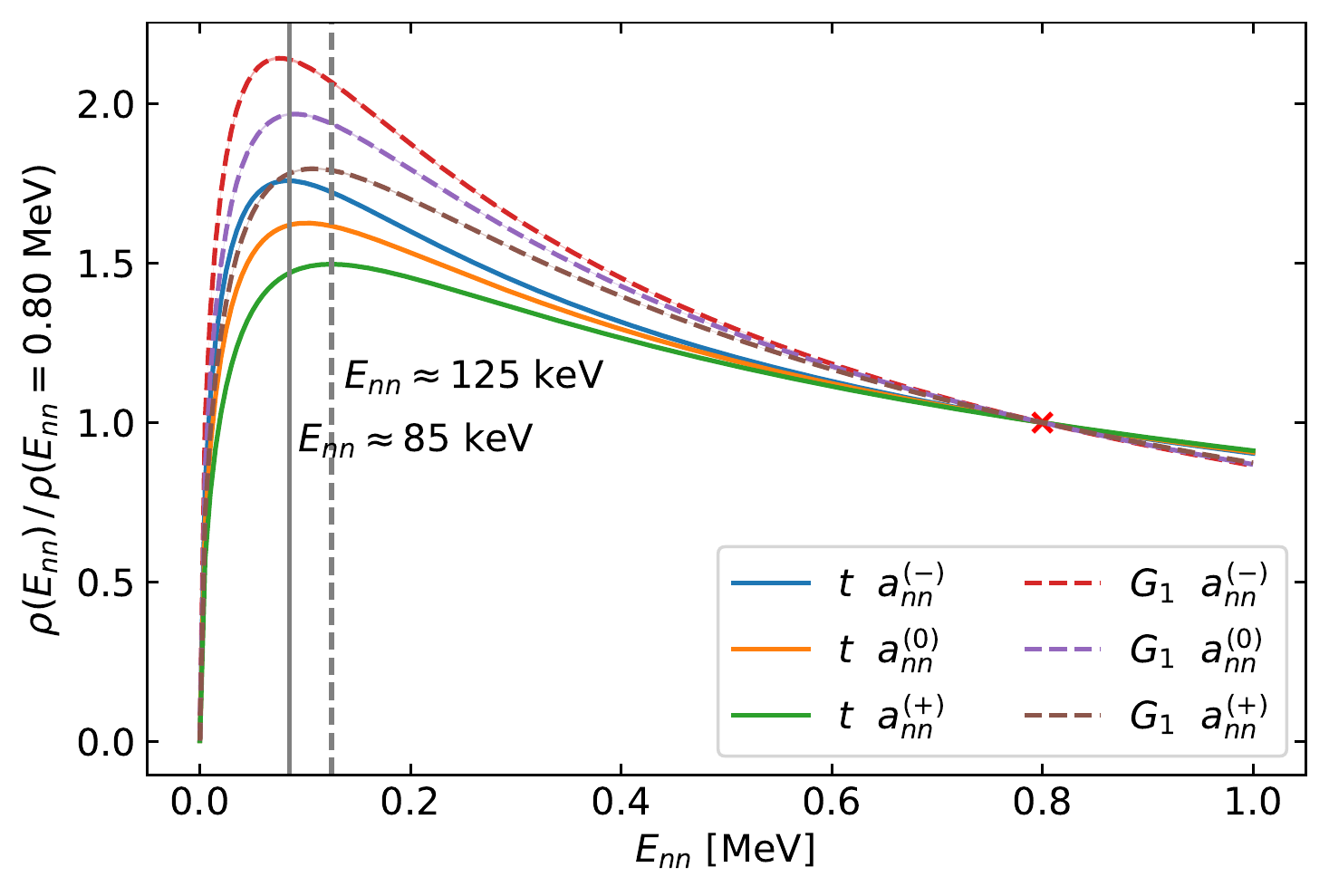}
      \caption{Comparison of \(nn\) relative-energy distributions for different \(nn\) scattering lengths obtained with different FSI schemes.
      The calculation using the \(nn\) t-matrix is labeled as `t'. \(r_{nn}=\SI{2.73}{\fm}\) is used.
      All results are computed using the projection \(\Psi_c{\K{p,q}}\) and 
      \lbh. Uncertainty bands based on comparison with calculation with
      half as many mesh points and \lbl~are negligible.
      In order to be independent of the normalization the distribution is divided by its value
      at the energy indicated by the red cross.
      The solid and dashed vertical lines indicate the approximate positions of the maxima in the t-matrix based FSI scheme
      for $a_{nn}^{(-)}$ and $a_{nn}^{(+)}$ respectively.}\label{fig:fsi_cmp}
    \end{figure}

    In \cref{fig:fsi_cmp} we compare 
    results for the \(nn\) relative-energy distribution obtained with the enhancement factor \(G_1\) and with the t-matrix
    treatment of FSI. We do this for three different \(nn\) scattering lengths, for which we use the following shorthand notation:
    \begin{equation}
      a_{nn}^{(+)} = \SI{-16.7}{\fm}\,, \qquad 
      a_{nn}^{(0)} = \SI{-18.7}{\fm}\,, \qquad
      a_{nn}^{(-)} = \SI{-20.7}{\fm}\,.
    \end{equation}

    As mentioned before, we do not calculate the absolute value of the distribution, but its shape.
    Therefore we normalize the distribution to a certain value at a certain position. Here we normalized to 1 at \(E_{nn} \approx 0.8\)~MeV.
    Not knowing the absolute value is no problem for determining the scattering length;
    the distribution will be fitted to the experimental data on the spectrum and the scattering length extracted from the shape.

    It can be seen that the scattering length has a significant influence on that shape. 
    When using distributions normalized to an arbitrary value at \(E_{nn} \approx \SI{0.8}{\mev}\), the main effect of the scattering
    length is to change the height of the peak located at relative energies of roughly \SI{100}{\kev}.
    Additionally, one can see that the two different procedures to include FSI yield curves of similar shape, but they are not quantitatively in agreement. 
    At a given \(a_{nn}\) the different FSI treatments produce different peak heights in the \(nn\) distribution. 
    The enhancement-factor  approach makes additional approximations beyond those involved when the $nn$ FSI is fully calculated from the $nn$ t-matrix. Therefore we trust
    the latter approach---with its full inclusion of the $nn$ FSI---more.

    We also calculated the distribution with the \(nn\) subsystem in the \(^3P_1\) partial wave by applying
    \cref{eq:psi_fsi} to the \(\Omega_c^{(1,1,1)}\) ground-state wave function component obtained with FaCE in
    setting F1.
    We found that this distribution is suppressed by a factor of at least 30 compared to the \(^1S_0\) distribution
    (in the \(E_{nn} < 1\)~MeV region).
    We compared the ground-state distributions as well and found that FSI increased  the suppression as anticipated at
    the beginning of \cref{sec:nn_gs_distrib}.

\section{\texorpdfstring{From the \(\boldsymbol{nn}\) relative-energy distribution to the \(\boldsymbol{nn}\) scattering length}{From the nn relative-energy distribution to the nn scattering length}}\label{sec:nn_scat_length}
\begin{figure}[H]
    \centering
    \includegraphics[width=.5\textwidth]{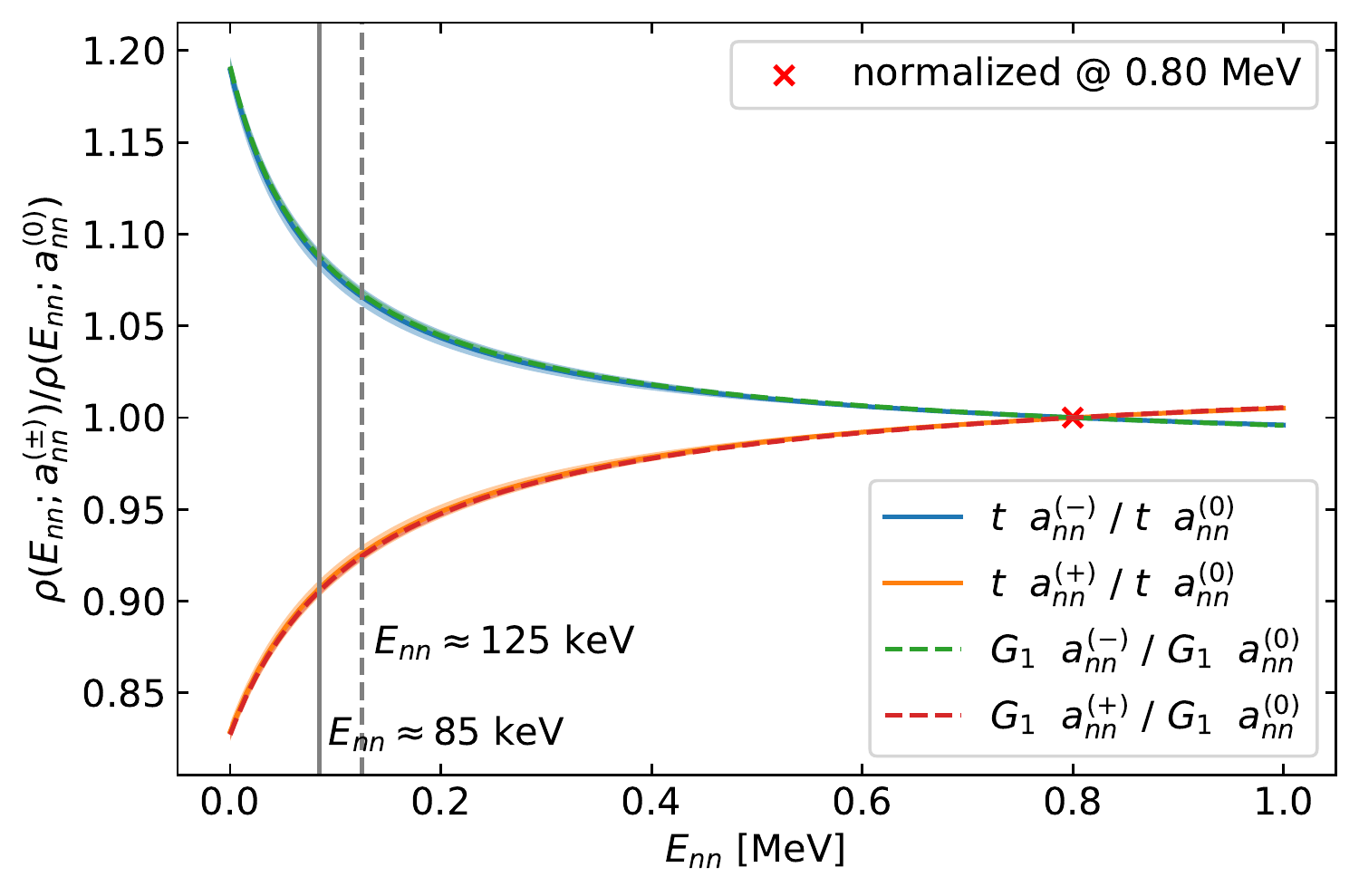}
    \caption{Ratios of \(nn\) relative energy distributions obtained with different scattering lengths for different FSI schemes in comparison.
    The calculation using the \(nn\) t-matrix is labeled as `t', \(r_{nn}=\SI{2.73}{\fm}\) is used.
    All results are based on \(\Psi_c{\K{p,q}}\).
    \lbh~was used. Uncertainty bands based on comparisons with calculation with
    half as many mesh points and \lbl~are shown.
    Due to a normalization of the distributions to a value of 1 at \(E_{nn}\approx \SI{0.8}{\mev}\), the quotients are
    at this point 1.
    The vertical lines indicate the approximate positions of the maxima in the t-matrix based FSI scheme
    for the lower and upper value of the scattering length.}\label{fig:fsi_cmp_rel}
  \end{figure}

  After showing results for the \(nn\) relative-energy distribution, we want to discuss in more detail how the scattering
  length can be extracted. Also, we want to discuss the role of the \(nn\) effective range.
  First, we quantify the influence of changing the scattering length by \SI{2}{\fm}.
  \Cref{fig:fsi_cmp_rel} shows quotients of the relative energy distributions obtained with different scattering lengths.

  It can be seen that, if our normalization
  scheme is used,  a change of the scattering length by \SI{2}{\fm} changes the peak heightby approximately 10\%.
  This change is almost completely independent of the method used to calculate the FSI.
  Additional calculations show that a change in the scattering length of \SI{0.2}{\fm} changes the peak height by
  about 1\%.

  In order to determine the scattering length at high precision it is necessary to know the influence of the \(nn\) 
  effective range on the distribution. 
  As with the scattering length, the effective range can enter the calculation at two stages.
  The first is the calculation of the ground state wave function of \hesix.
  At this level, the influence of the scattering length is low.
  As the scattering length is a leading-order parameter and the effective range is a next-to-leading-order parameter, we
  expect its influence on the \hesix~wave function to be very small.
  The second stage is taking the FSI into account. 
  In this step, the scattering length plays a significant role.
  Therefore, we cannot exclude a non-negligible influence of the range in this step.
  We investigate the influence of the effective range in the t-matrix based FSI approach and in the approach employing the 
  enhancement factor \(G_1\) by using the following values:
  \begin{equation}
    r_{nn}^{(+)} = \SI{3.0}{\fm}\,,  \qquad 
    r_{nn}^{(0)} = \SI{2.73}{\fm}\,, \qquad
    r_{nn}^{(-)} = \SI{2.0}{\fm}\,,
  \end{equation}
  which is a rather large variation around the common literature value of \(r_{np}({}^1S_0) = \SI{2.73 \pm 0.03}{\fm} \approx r_{nn}\) \cite{preston1975structure}.
  Note, that we included \(r_{nn}\) only in the calculation of the FSI but not in the calculation of the ground-state wave function,
  as its influence there should be negligible.
  While the effective range is varied from \(r_{nn}^{(-)}\) to \(r_{nn}^{(+)}\) the change in the distribution is small:
   less than 1\% at peak position.
  Details can be seen from \cref{fig:fsi_rnn} in \cref{ap:add_plots}.

  As a conclusion, these results show that scattering length has a significant influence on the
  \(nn\) relative energy distribution and that the effective range does not.
  Thus, the distribution is suitable for extracting the scattering length.

\section{Outlook}\label{sec:outlook}

Already at its present accuracy our calculation will be able to provide a precise determination of the  \(nn\) scattering length using data from the measurement of the \({}^6{\rm He}(p,p\alpha)nn\) reaction that has been approved 
at RIKEN~\cite{nn_scat_len_ribf_prop2018}.
We also plan to  increase the accuracy
of the EFT calculation, i.e., make the uncertainty band narrower.
First, we will move to NLO Halo EFT calculations of the ground-state wave function.
While at \(E_{nn}=\SI{1}{\mev}\) the LO uncertainty is approximately 20\%,
the NLO uncertainty at this position will be around 5\%.
Second, we also want to improve the treatment of the final-state interaction in EFT. 
For this purpose we will develop an EFT framework for knockout reactions like the one considered here.
This will enable the inclusion of both the \(n\alpha\) interaction after knockout and corrections to the assumption that the \(\alpha\) particle removal does not affect the \(nn\) relative-energy distribution---or at least it will allow us to constrain such effects
as occurring at a high order in a small expansion parameter. 
The resulting EFT approach to the entire \({}^6{\rm He}(p,p\alpha)nn\) reaction will allow us to rigorously assess the full uncertainty of the two-step calculation
we have carried out here.

The reaction \(t(p,2p)2n\) would also facilitate a measurement of the \(nn\) scattering length along the same lines as those discussed in this paper.
This would be a valuable reaction to examine: using two different nuclei as neutron sources checks the reliability of the
result for the scattering length. 
As in the case of \hesix~the ground state wave function of the triton can be calculated in an EFT. In the case of
the triton it is the well-established pionless EFT, in which the neutron and proton are the low-energy degrees of freedom~\cite{Bedaque:1999ve}.
This EFT has the advantage that there is no relevant \(p\)-wave interaction, but the triton 
has a two-neutron separation energy of \SI{8.48}{\mev} and so is more strongly bound than \hesix. The treatment of 
its proton-induced breakup would thus involve 
a bigger expansion parameter and larger uncertainties at the same order than in the case of the EFT for \hesix.
This, though, is compensated by the fact that pionless EFT for the triton is established
up to N\(^2\)LO~\cite{Bedaque:2002yg}.

\begin{acknowledgments}
T.A., H.-W.H. and M.G. acknowledge support by the Deutsche Forschungsgemeinschaft (DFG, German Research Foundation) - Project-ID 279384907 - SFB 1245.
T.F. was a Fulbright Scholar visiting the University of Iowa at Ames when this work started and thanks for the hospitality received,
 and also thanks Conselho Nacional de Desenvolvimento Científico e Tecnológico (CNPq) under Grant no. 308486/2015-3,  Fundação de Amparo à Pesquisa do Estado de São Paulo (FAPESP) under the thematic projects  2017/05660-0, INCT-FNA project 464898/2014-5 and Coordenação de Aperfeiçoamento de Pessoal de Nível Superior—Brazil (CAPES—Finance Code 001).
C.A.B. is partially supported by the U.S. Department of Energy Grant No. DE-FG02-08ER41533 and funding contributed by the LANL Collaborative Research Program by the Texas A\&M System National Laboratory Office and Los Alamos National Laboratory.
D.R.P. acknowledges support from the U.S. Department of Energy (contract DE-FG02-93ER40756) and the ExtreMe Matter Institute.
\end{acknowledgments}

\appendix

\section{Calculation of wave functions in Halo EFT}\label{ap:wf_heft}
Here we describe, how wave functions of the type
\begin{equation}
  \Psi_i{\K{\pq}} \coloneqq \ibraket{i}{p,q;\Omega_i}{\Psi}{}
\end{equation}
can be calculated starting from the \(F_i{\K{q}}\) that are the solutions of the integral equations obtained using the Faddeev formalism in our Halo EFT approach.
The procedure is not specific to Halo EFT, it is a general procedure for Faddeev equations in momentum space.
However, in the course of this description we use identities specific to \hesix, mainly affecting the partial-wave states.
In this context, \(\Omega_i\) is the multi-index specifying the quantum numbers of the three-body system with particle \(i\) as a spectator under the condition
that the total quantum numbers are the ones of the \hesix~ground state and the numbers for \(jk\) subsystem characterize the interaction
channel.
The basics of the Faddeev equations are described, e.g., in Ref. \cite{gloeckle83}.
Additionally, we use results and notation of Ref. \cite{Gobel:2019jba}.

Making use of the decomposition of the total state \(\ket{\Psi}\) described in \cref{eq:psi_f_i}, we can write
\begin{equation}
  \Psi_i{\K{\pq}} \coloneqq \ibraket{i}{p,q;\Omega_i}{\Psi}{} = \sum_j \ibraket{i}{p,q;\Omega_i}{\psi_j}{}\,,
\end{equation}
whereby \(\ket{\psi_i} \coloneqq G_0 t_i \ket{F_i}\) holds.
In an intermediate step, we calculate
\begin{equation}
  \psi_i{\K{\pq}} \coloneqq  \ibraket{i}{p,q;\Omega_i}{\psi_i}{} = 4\pi G_0^{\K{i}}{\K{p,q;-\btz}} \reg{i}{p} \tau_i{\K{q;-\btz}} F_i{\K{q}} \,,
\end{equation}
where \cref{eq:rep_f_i,eq:t_3bd,eq:t_2bd} as well as the definition \(\tau_i{\K{q;E_3}} \coloneqq \tau_{jk}{\K{E_3 - \frac{q^2}{2\mu_{i(jk)}} }}\)
were used.
Additionally, \(G_0^{(i)}{\K{p,q;E_3}} \coloneqq \K{E_3 - p^2 / \K{2\mu_{jk}} - q^2 / \K{2\mu_{i(jk)}} }^{-1}\) holds.
Consequently we have a relation between \(\psi_i{\K{\pq}}\) and the numerically determined \(F_i{\K{q}}\).
We use this result to continue the calculation of the wave function of the total state\footnote{
  Note that there might be additional non-vanishing wave functions \(\Psi_i^{(\Omega)}{\K{\pq}} \coloneqq  \ibraket{i}{p,q;\Omega}{\Psi}{}\)
  where \(\Omega\) is a fixed multi-index not contained in the set of the \(\Omega_i\), which are the quantum numbers of the interaction channels.
  These \(\Psi_i^{(\Omega)}{\K{\pq}}\) are calculated as described in \cref{eq:Psi_i_from_F_i}
  with the modification that the \(\Omega_i\) (not the \(\Omega_j\)) has to be replaced by the \(\Omega\) of interest.
  Since this multi-index \(\Omega\) has no index naming him, we would call \(f_{ij}\) now \(f_{\Omega ij}\).
  (The \(\kappa\) functions are independent of \(\Omega\).)
  We calculated such wave functions, e.g., a \(\Psi_c^{(\Omega)}{\K{\pq}}\) for \(\Omega = \Omega_c^{(0,2,2)}\) and one for
  \(\Omega = \Omega_c^{(1,1,1)}\) (notation from \cref{sec:tb_calcs}).
}:
\begin{align}\label{eq:Psi_i_from_F_i}
  \Psi_i{\K{\pq}} &= \sum_j \ibraket{i}{p,q;\Omega_i}{\psi_j}{} 
  = \sum_j \rint{\pp} \rint{\qp} \underbrace{\ibraket{i}{p,q;\Omega_i}{\pp,\qp;\Omega_j}{j}}_{\mathclap{
    = \int \dd x f_{ij}{\K{p,q,x}} \de{\pp - \kappa_{ijp}{\K{p,q,x}}} \de{\qp - \kappa_{ijq}{\K{p,q,x}}} / \K{p'^2 q'^2}
  }} \psi_j{\K{\pp,\qp}} \nonumber \\
&= \sum_j \int \dd x f_{ij}{\K{\pq,x}} \psi_j{\K{ \kappa_{ijp}{\K{p,q,x}}, \kappa_{ijq}{\K{p,q,x}} }} \,,
\end{align}
where we used \(\ibraket{j}{p,q;\Omega}{\psi_j}{} = \kd{\Omega}{\Omega_j} \psi_j{\K{\pq}} \), which follows from the properties of
the used t-matrices
and denoted \(\cos{\thpq}\) as \(x\).
For the momenta \(\kappa_{ijk}\) (\(k \in \{p,q\}\)) we use the notation of Appendix B.1 from Ref. \cite{Gobel:2019jba}.
Formulas for the overlaps and therefore implicitly also for the \(f_{ij}{\K{\pq,x}}\) are given in Appendix B.4 of that reference.
We evaluate the angular integral numerically, the formula is based on simplifications described
in Appendix B.2 of Ref. \cite{Gobel:2019jba}.
Note that the antisymmetrization under \(nn\) permutation causes some complications, but is just a special case of the more general 
structure described before.

The specific expression for \(\Psi_c{\K{p,q}}\) we are using then reads
\begin{align}
  \Psi_c\K{p,q}
  &= 2\pi \int_{-1}^{1} \dd{\cos{\thpq}} \K{ a_c \frac{\sqrt{2}}{4\pi}\, \K{ \kcnph \cdot \kcnqh } + \tilde{a}_c \frac{\sqrt{2}}{4\pi} \K{ \kcnpph \cdot \kcnqph } + \frac{d_c}{4\pi}} \,.
\end{align}
We use the following definitions:
\begin{align}\label{eq:pbd_coeff_c_first}
  a_c          &\coloneqq \psi_n{\K{\kcnp, \kcnq}} \,, 
  & \tilde{a}_c  &\coloneqq \psi_n{\K{\kcnp^\prime, \kcnq^\prime}} \,,
  & d_c          &\coloneqq \psi_c{\K{p,q}} \,.
\end{align}
Note that the \(\kcnph\), \(\kcnqh\), \(\kcnpph\), \(\kcnqph\) and thereby also the ``coefficients" \(a\), \(\tilde{a}\), \(d\) depend on
the momenta \(p\), \(q\) and in general also on \(x=\cos{\thpq} \,.\) 
E.g., \(a_c = \psi_n{\K{ \kcnp{\K{p,q,x}}, \kcnq{\K{p,q,x}} }}\) holds.

\section{\texorpdfstring{Influence of the \(\boldsymbol{nn}\) effective range on the \(\boldsymbol{nn}\) relative-energy distribution}{Influence of the nn effective range on the nn relative-energy distribution}}\label{ap:add_plots}
\begin{figure}[H]
    \centering
    \includegraphics[width=.5\textwidth]{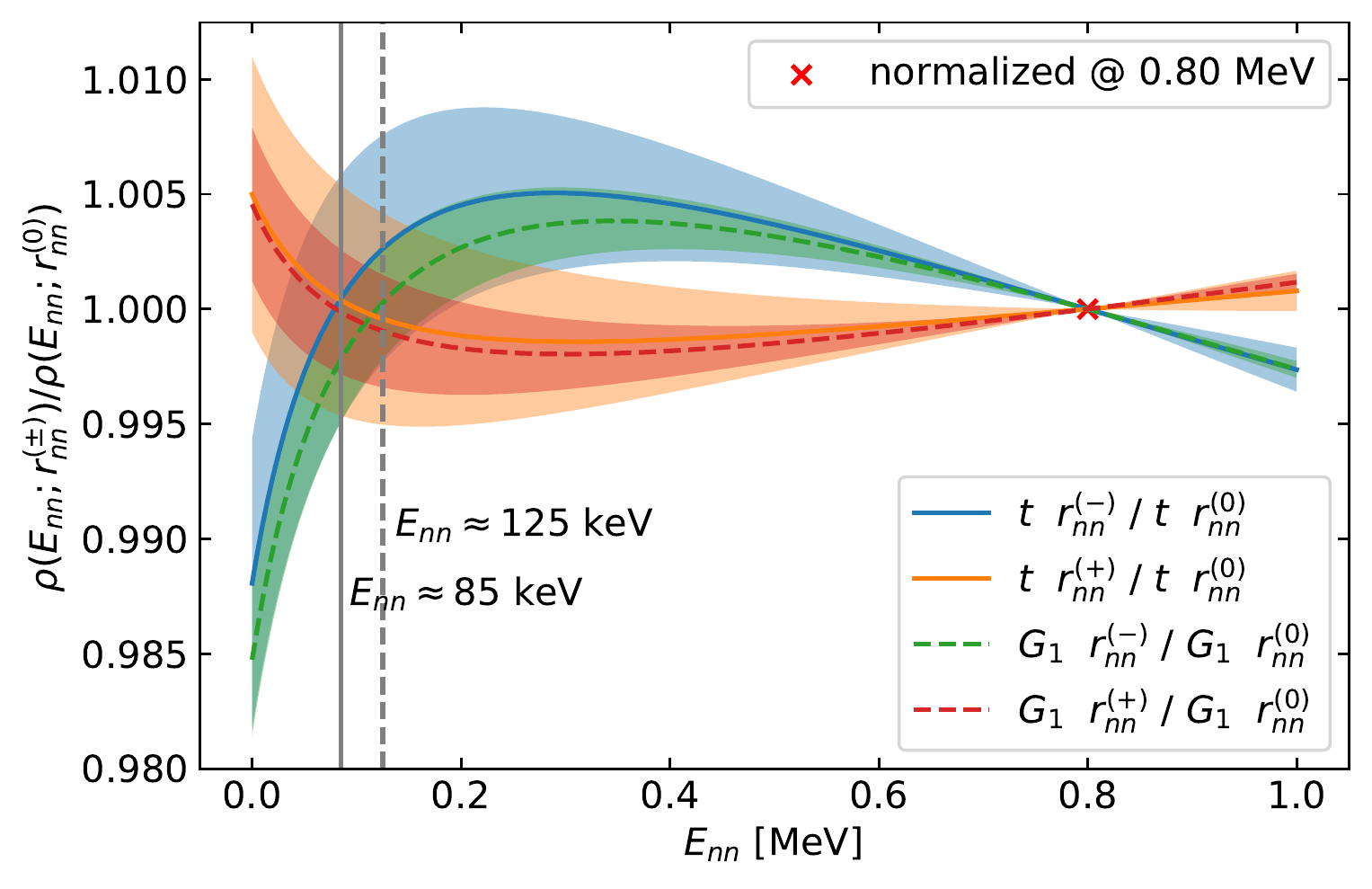}
    \caption{
      Ratios of \(nn\) relative energy distributions obtained with different effective ranges
      for different FSI schemes in comparison.
      The definitions \(r_{nn}^{(+)} = \SI{3.0}{\fm}\),  \(r_{nn}^{(0)} = \SI{2.73}{\fm}\) and
      \(r_{nn}^{(-)} = \SI{2.0}{\fm}\) hold. All results are based on \(\Psi_c{\K{p,q}}\).
      \lbh~was used. Uncertainty bands based on comparisons with calculation with
      half as many mesh points and \lbl~are shown.
      Due to a normalization of the distributions to a value of 1 at \(E_{nn}\approx \SI{0.8}{\mev}\), the quotients are
      at this point 1.
      The vertical lines indicate the approximate positions of the maxima in the t-matrix based FSI scheme
      for the lower and upper value of the scattering length. 
    }\label{fig:fsi_rnn}
  \end{figure}

  For the planned experiment, also the dependency of the \(nn\) relative-energy distribution on the \(nn\) effective range is relevant.
  Ideally, this dependency would be small in order not to complicate extraction of the scattering length from the measured spectrum.
  \Cref{fig:fsi_rnn} shows ratios of final-state \(nn\) relative-energy distributions obtained with different values for the \(nn\) effective range. The influence of the effective range variations at the level of the ground-state was neglected in these calculations, up to this point only \(r_{nn}=2.73\) fm was used there. What is shown is therefore the effect on the FSI when \(r_{nn}\) is varied by 1~fm. 

  The overall variation of the effective range by \SI{1}{\fm} has only a small influence on the distribution.
  The changes caused by this variation are less than 1\% at the peak position.
  The bands showing the numerical uncertainty may appear large, but their absolute size is of the same order
  as in \cref{fig:fsi_cmp_rel}.
  However, they cover large parts of the plot 
  because of the small influence of the effective range.

  Finally, we explain how these numerical uncertainties of ratios of distributions were estimated
  in \cref{fig:fsi_cmp_rel} and \cref{fig:fsi_rnn}:
  The numerical uncertainty of the ratio \(r\) of distributions \(\rho^{(1)}\) and \(\rho^{(2)}\)
  given by
  \begin{equation}
    r{\K{E_{nn}}} \coloneqq \rho^{(1)}{\K{E_{nn}}} / \rho^{(2)}{\K{E_{nn}}} 
  \end{equation}
  was estimated according to 
  \begin{equation}
    \Delta r{\K{E_{nn}}} = \sqrt{ \K{ \frac{ \Delta \rho^{(1)}{\K{E_{nn}}} }{ \rho^{(2)}{\K{E_{nn}}} } }^2 +
    \K{ \frac{ -\rho^{(1)}{\K{E_{nn}}} }{ \K{ \rho^{(2)}{\K{E_{nn}}} }^2 } \Delta \rho^{(2)}{\K{E_{nn}}} }^2 }
  \end{equation}
  by using the uncertainties of the distributions denoted by \(\Delta \rho^{(1)}\) and \(\Delta \rho^{(2)}\).
  That is the standard formula for the propagation of uncertainties based on the linearization of the functions
  under the assumption that the two distributions are not correlated.
  If we would assume a correlation of 1 between the distributions (for all energies), the uncertainty bands would get much smaller.
  While this might be a reasonable approximation, we chose to draw more pessimistic uncertainty bands
  by not using it.

\bibliographystyle{apsrev4-2}
\bibliography{he6_and_nn_scattering_length.bbl}

\end{document}


\title{\texorpdfstring{Supplemental material for\\``Neutron-neutron scattering length from the $\boldsymbol{^6}$He$\boldsymbol{(p,p\alpha)nn}$ reaction"}{
Supplemental material for "Neutron-neutron scattering length from the He-6(p,p alpha)nn reaction"
}}

\author{Matthias G\"{o}bel \orcid{0000-0002-7232-0033}}

\author{Thomas Aumann \orcid{0000-0003-1660-9294}}

\author{Carlos A. Bertulani \orcid{0000-0002-4065-6237}}

\author{\\Tobias Frederico \orcid{0000-0002-5497-5490}}

\author{Hans-Werner Hammer \orcid{0000-0002-2318-0644}}

\author{Daniel R. Phillips \orcid{0000-0003-1596-9087}}

\keywords{Neutron scattering length, knockout reactions, indirect techniques}

\date{June 1, 2021}

\begin{abstract}
In this supplemental material, we provide additional details regarding some of our findings 
and give additional information on our calculations, especially on the model calculations.
In \cref{sec:diff_analysis} the differences between the leading-order Halo EFT calculation and the model
calculations are comprehensively analyzed.
\Cref{sec:gs_distrib_nn_sl} contains a plot showing the small influence of the \(nn\) scattering length on the
ground-state \(nn\) relative-energy distribution.
The two-potential scattering theory used in the main paper is briefly summarized in \cref{sec:two_pot_scat_theory}.
\Cref{sec:FSI_EFs} discusses the different available FSI enhancement factors and a ``continuous FSI enhancement factor''
is introduced.
The relations used for computing the FSI effects are laid out in \cref{ap:fsi_formula}.
\Cref{ap:face_mom_sp_wf} and \cref{ap:turn_poh_off} give more information on the model calculations based on 
the computer code FaCE.
Finally, \cref{ap:face_num_param} lists the computational parameters of these model calculations.
\end{abstract}

\maketitle

\section{\texorpdfstring{Ground-state \(\boldsymbol{nn}\) relative-momentum distributions: Analysis of discrepancies}{Ground-state nn relative-momentum distributions: Analysis of discrepancies}}\label{sec:diff_analysis}

The comparison of the EFT result with the \lgmacronym~result in \hcref{Fig. 2 of the main part} has clearly shown that the results are consistent:
the \lgmacronym~result lies within the uncertainty band of the leading-order Halo EFT calculation.
Especially in the region of \(p < \SI{30}{\mev}\) relevant for the proposed experiment the agreement is good.
Nonetheless, we are interested in the sources of the discrepancies and in the possible results we can expect from
a Halo EFT calculation at next-to-leading-order.

This analysis of discrepancies is already summarized in \hcref{Section III.A of the main part}.
The purpose of this supplement is to give more detailed information.
In the first step of our analysis we list the most obvious causes of the difference:
\begin{enumerate}
  \item phenomenological \lgmacronym~three-body potential;
  \item different effective range expansion (ERE) parameters;
  \item different off-shell properties of the two-body interactions, which are possibly not completely
        compensated by the used three-body forces (see Ref. \cite{Polyzou:1990} for details on this topic);
  \item EFT calculation is LO; at higher order:
    \begin{enumerate}
      \item additional terms of the ERE can change the results (especially the unitary term of the \(nc\) system);
      \item interactions in additional partial waves such as \(^2P_{1/2}\) in the \(nc\) system contribute.
    \end{enumerate}
\end{enumerate}

In order to estimate which of the possible origins are more relevant we do some additional calculations.
We introduce variations of our default physical input parameters for \lgmacronym. The parameter sets are
\begin{itemize}
  \item \lgmacronym 1: standard version of \lgmacronym~input parameters for \hesix;
  \item \lgmacronym 2: \lgmacronym 1, but \(nc\) interaction only in \(^2P_{3/2}\) (\(nc\) interaction in \swave, \dwave~and \(^2P_{1/2}\) is turned off);
  \item \lgmacronym 3: \lgmacronym 2, but three-body potential deactivated.
\end{itemize}
Physically speaking, we have introduced alternative Hamilton operators for the \lgmname~to be solved by FaCE.
With the variations of the \lgmacronym~calculation we can analyze the influence of the \(nc\) channels which are not present in the leading-order
Halo EFT calculation. And we can investigate the influence of the phenomenological three-body potential.

While turning the \(nc\) \swave~ and \dwave~interaction off is easy, turning the interaction in \(^2P_{1/2}\) off and keeping the one
in \(^2P_{3/2}\) unchanged is a bit trickier. We do it by tuning the depth of the central and of the spin-orbit potential.
This achieves exactly the effect we want, since both potentials have the same range.
Details can be found in \cref{ap:turn_poh_off}.

Since the used three-body potential is phenomenological, its parameters are
not directly fixed by more fundamental processes. Thus we employ also variations of \lgmacronym 1 and \lgmacronym 2 where the range of three-body potential is reduced.
For these the range parameter \(\rho_\mathrm{3B}\) a value of \SI{2.5}{\fm} instead of \SI{5.0}{\fm} was used, and they are 
denoted by \lgmacronym 1SR and \lgmacronym 2SR. In all variations of 
the \lgmacronym~calculation---except for \lgmacronym 3---the depth of the three-body potential is adjusted to produce the correct binding energy for the three-body
system\footnote{  
  In order to do this we used a root-finding routine for running FacE and adjusting
  the three-body force to the desired binding energy.
}.

In order to be able to do more fine-grained comparison we use another model, which is less elaborate
than those that can be solved by FaCE, but is also useful for this analysis.
We turn our Halo EFT code, which solves the Faddeev equations in momentum space, into a model calculation by using a different set
of t-matrices. Separable potentials with Yamaguchi form factors are able to provide reasonable fits to \(nn\) and \(nc\) phase shifts.
We use the separable t-matrices corresponding to those potentials and call this model ``Yamaguchi model" (YM).
It is similar to calculations in Refs. \cite{Hebach:1967bpg,Shah:1970wu,Ghovanlou:1974zza}.
Since it is a momentum-space calculation, it is also easy to use the Halo EFT type three-body force in this model.
Here, the t-matrices are connected to the potential parameters via the Lippmann-Schwinger equation.
The potential parameters are in turn fixed by the lowest-order parameters of the effective range expansion.
The Yamaguchi separable potential in partial wave \(\tilde{l}\) reads
\begin{equation}\label{eq:V_YG}
    \mel{p,l}{V_{\mathrm{YG};\tilde{l}}}{\pp,l^\prime} = 4\pi \kd{l}{l^\prime} \kd{l}{\tilde{l}} 
    g_{\mathrm{YG};l}{\K{p}} \lambda_l g_{\mathrm{YG};l}{\K{\pp}} \,,
\end{equation}
where \(g_{\mathrm{YG};l}{\K{p}} = p^l \K{1 + \frac{p^2}{\beta_l^2}}^{-l-1}\) is the Yamaguchi form factor.
This potential thus has two parameters: the strength parameter \(\lambda_l\) and the regulation scale \(\beta_l\).
The corresponding t-matrix\footnote{
  Note, that this is the general formula for the two-body t-matrix of this Yamaguchi model.
  For the three-body calculation, the three-body embeddings of the t-matrices are necessary.
  Such an embedding is described by \hcref{Eq. (16) of the main part}.
}
\begin{equation}
    \mel{p,l}{ t_{\mathrm{YG};\tilde{l}}{\K{E}} }{\pp,l^\prime} = 4\pi \kd{l}{l^\prime} \kd{l}{\tilde{l}}
    g_{\mathrm{YG};l}{\K{p}} \tau_l{\K{E}} g_{\mathrm{YG};l}{\K{\pp}} \,,
\end{equation}
is given by the Lippmann-Schwinger equation for separable potentials:
\begin{equation}\label{eq:lse_sep}
    \tau_l^{-1}{\K{E}} = \lambda_l^{-1} - 4\pi \rint{\q} \frac{ g_{\mathrm{YG};l}^2{\K{q}} }{ E - q^2/\K{2\mu} + \ci \epsilon} \,.
\end{equation}
The real part of \(\tau_l^{-1}{\K{E}}\) can be expanded in powers of
\(k^2\) to match it with the effective range expansion. The strength of the potential and
the regulation scale are then determined by the lowest two effective-range-expansion parameters.
In our standard YM calculation we use the following parameters:
\(\lambda_{nn}= \SI{-4.293e-07}{\mev^{-2}}\), \(\beta_{nn}= \SI{230.1}{\mev}\), \(\lambda_{nc}= \SI{-1.929e-11}{\mev^{-4}}\) and \(\beta_{nc}= \SI{296.2}{\mev}\).
As in the Halo EFT calculation the \(nn\) interaction is purely \swave, while the \(nc\) interaction is purely in \(^2P_{3/2}\).
(Note that for the sake of simplicity we omitted the indices for total subsystem angular momentum \(j\) in \cref{eq:V_YG}
as well as in the equations for the t-matrix elements, even though the interaction in \(p\)-waves is dependent on \(j\).) 
In order to investigate certain effects, it is also possible to remove certain terms from the effective-range expansion.
We perform also the following modified YM calculations differing in the \(nc\) t-matrices:
\begin{itemize}
\item  The standard one denoted by YM1 contains terms up to \(k^8\) in the denominator of the \(nc\) t-matrix
as they follow from \cref{eq:lse_sep} (higher-order terms do not exist). 
\item In contrast, in the so-called YM2 calculation the unitary term is missing from the \(nc\) t-matrix, i.e., there is no \(i k^3\) in \(\tau_l^{-1}{\K{E}}\).
This is done in order to have a version more similar to the leading-order Halo EFT calculation, where the unitarity term of the \(nc\) 
interaction is not included. (The unitarity term is an NLO effect in the power counting devised in Ref.~\cite{Bedaque:2003wa} and used in Ref.~\cite{Ji:2014wta}.) 
\item The YM3 model tries to be even closer to the LO Halo EFT \(nc\) amplitude by removing the \(k^4\) and the \(k^6\) term from  \(\tau_l^{-1}{\K{E}}\).
The \(k^8\) term is not removed in order to avoid unphysical poles
in the probed region of the t-matrices \(k^2 \leq - 2\mu \btz\).
\end{itemize}

Before we compare the results of the described calculations, we make some general remarks
that we want to keep in mind during the analysis.
Firstly, potentials are not observable.
So, when we discuss the actions of different potentials, we talk about the properties of tools but not about physical observables.
Secondly, when doing various calculations differing in the used potentials, one has to be aware of the fact
that the action of a potential in terms of yielding a specific wave function is not absolute but
relative to the other potentials in play.
Thirdly, we want to consider the statements of Ref. \cite{Polyzou:1990} on the relation between two- and three-body forces.
Polyzou and Gl\"ockle found that different sets of two-body interactions with the same bound states and phase shifts
can be made to yield the same three-body bound states and three-body phase shifts, if the appropriate three-body force is used
together with one of the two two-body interactions.
That means that is pointless to discuss whether a three-body force is present in a system or not.
As they state, however, it can be meaningful to discuss whether a model needs a three-body force.
We will focus on this aspect, and examine whether certain three-body forces are able to compensate for the different off-shell properties of two-body
potentials in \lgmacronym, LO Halo EFT, and the YM. 
Even though the local \(l\)-dependent potentials of the \lgmacronym~calculation and the t-matrices in the EFT calculation yield approximately
the same phase shifts at low energies, they do not have the same off-shell behavior.

\begin{figure}[H]
    \centering
    \includegraphics[width=.42\textwidth]{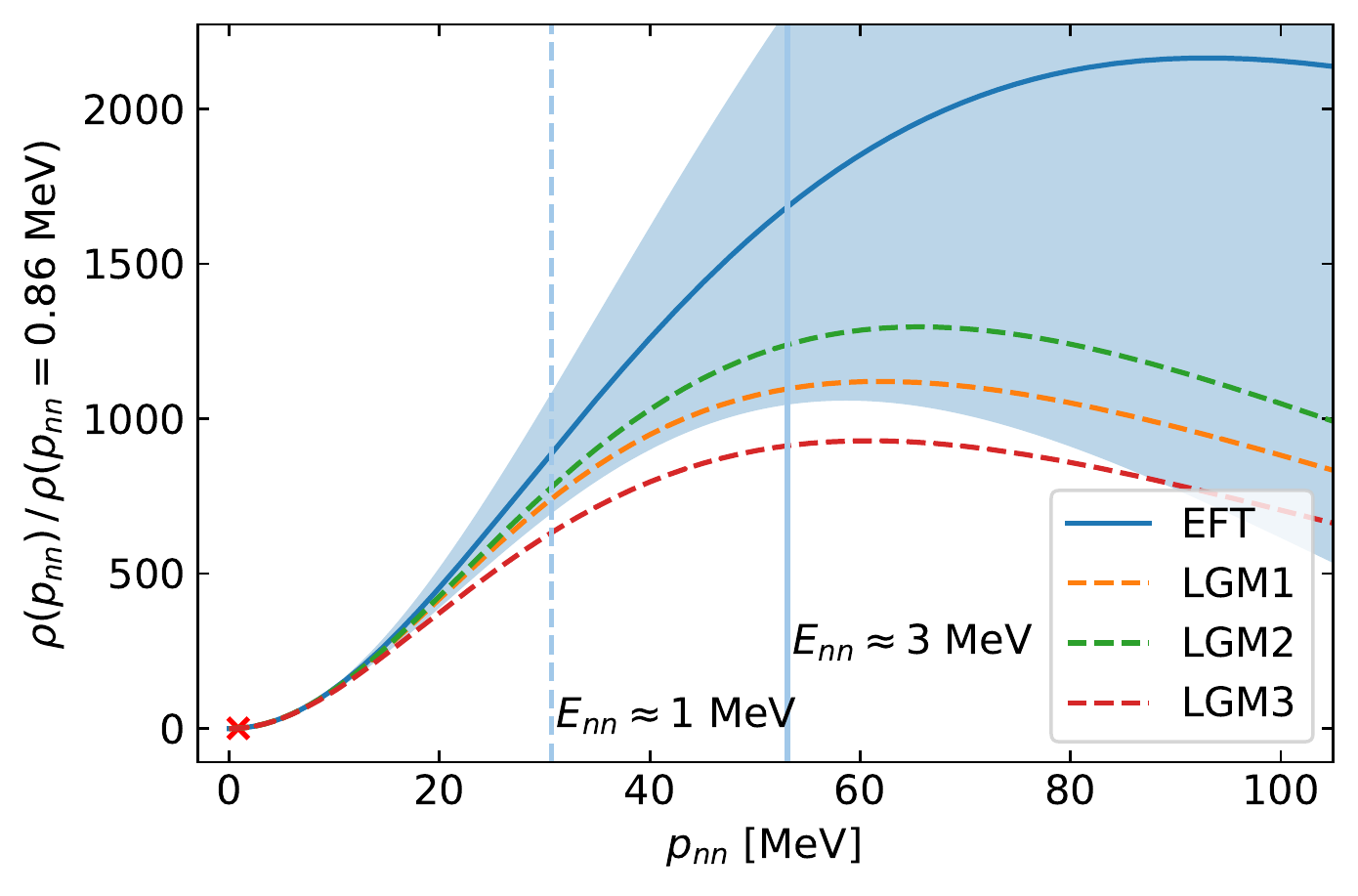}
    \includegraphics[width=.42\textwidth]{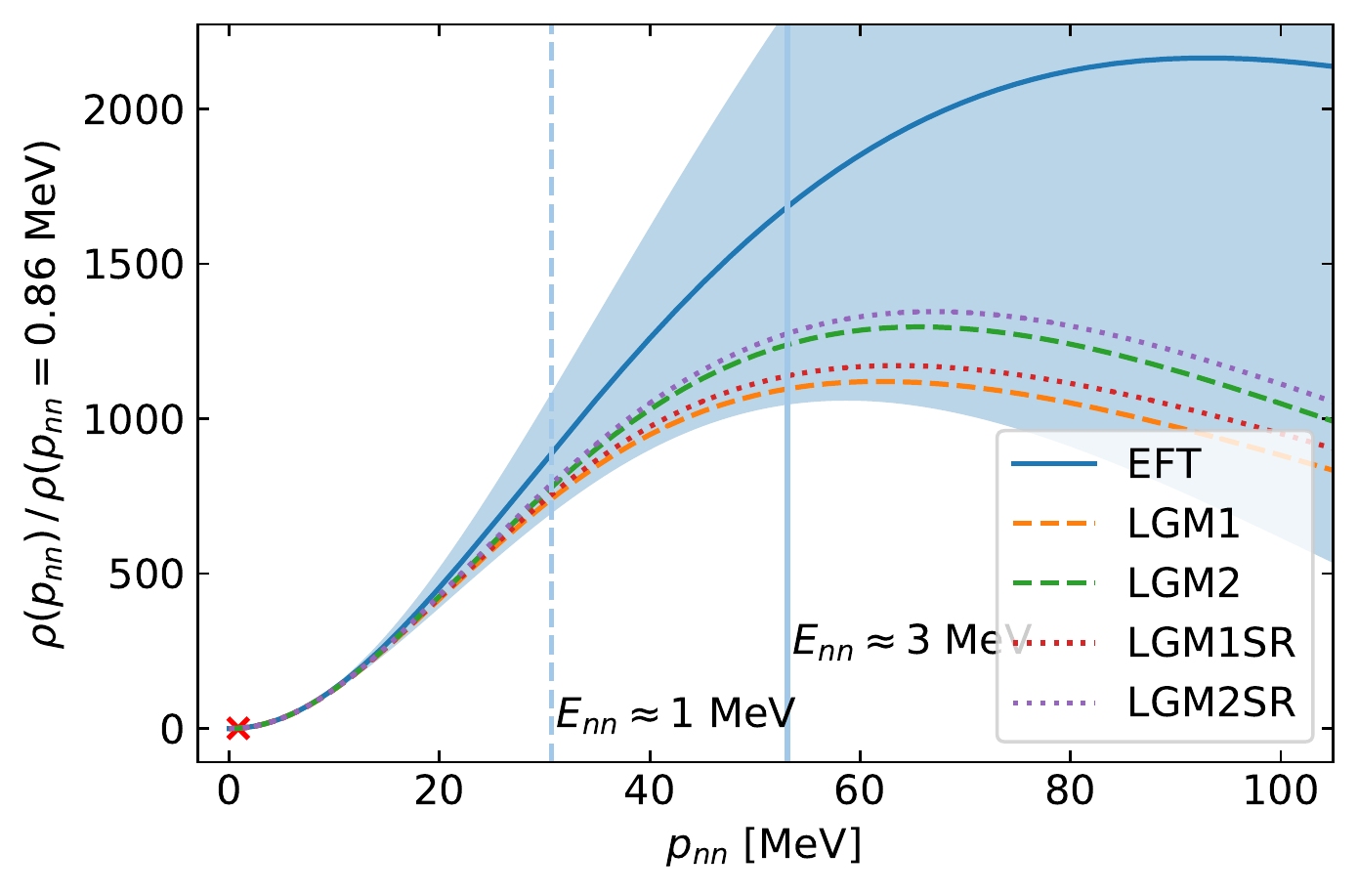}
    \caption{\lgmacronym~results in comparison with the Halo EFT result. 
    The depth parameter of the three-body potential in the \lgmacronym~calculations \lgmacronym 1 and \lgmacronym 2 was in both cases tuned to reproduce \(\btz\). 
    \lgmacronym 1SR and \lgmacronym 2SR use a more short-ranged version of this three-body potential, the depth parameter was in both cases readjusted to
    the physical binding energy.
    Again, the distributions are divided by their value at a
    certain position, which is indicated by the red cross.
    The vertical lines indicate relative energies of \SI{1}{\mev} and \SI{3}{\mev}.}\label{fig:boris2}
\end{figure}

\Cref{fig:boris2} shows in the left panel results of \lgmacronym 1, \lgmacronym 2 and \lgmacronym 3 in comparison with the EFT result.
In the right panel the same comparison is shown including \lgmacronym~calculations with more short-ranged three-body forces.
It can be seen that \lgmacronym 2, having only the interaction channels present in the LO Halo EFT calculation,
agrees better with the EFT result than \lgmacronym 1.
Since \lgmacronym 2 is in terms of its two-body interactions more similar to the LO EFT calculation than \lgmacronym 1 and the three-body forces
are tuned to the binding energy, this agrees with our expectation.
If the phenomenological three-body potential of the \lgmname~is turned off (\lgmacronym 3), the agreement gets worse.
Having Ref. \cite{Polyzou:1990} in mind, we would conclude the combination of the EFT three-body force in the EFT calculation
and the phenomenological three-body force of \lgmacronym~is able to compensate at least partly for the differences in the off-shell
behavior of the EFT and \lgmacronym~potentials. 
Furthermore, looking at the right panel, we observe that using the more short-ranged three-body potential
in the \lgmname s improves the agreement slightly.
However, the influence of this variation seems to be small:
 although \lgmacronym 2 is closer to the LO Halo EFT result than \lgmacronym 1 it is still some distance away. Will this gap be reduced at NLO?

We want to remind the reader that there is no problem, Halo EFT at LO and \lgmacronym~agree within the uncertainty bands.
But we now analyze this further as we want to get a more detailed understanding of the difference.
One relevant not yet investigated source of possible deviations is the fact that the power counting we use here retains only the 
lowest terms
of the effective range expansion in the 
\(nn\) t-matrix in \(^1S_0\) and the \(nc\) one in \(^2P_{3/2}\). Furthermore, while in case of the \(nn\) t-matrix the unitary term is contained, that is not true
for the \(nc\) t-matrix.
In order to investigate the influence of these truncations of the two-body t-matrices we do the Yamaguchi model calculation.
While the standard version denoted by YM1 contains the full \(nc\) t-matrix according to the Lippmann-Schwinger equation for
separable potentials, we do variations with the unitary term missing (YM2 and YM3).
All three Yamaguchi models contain a three-body force of the same form as that used in the Halo EFT calculation.
As usual, its strength is tuned to the three-body binding energy.

\begin{figure}[H]
    \centering
    \includegraphics[width=.42\textwidth]{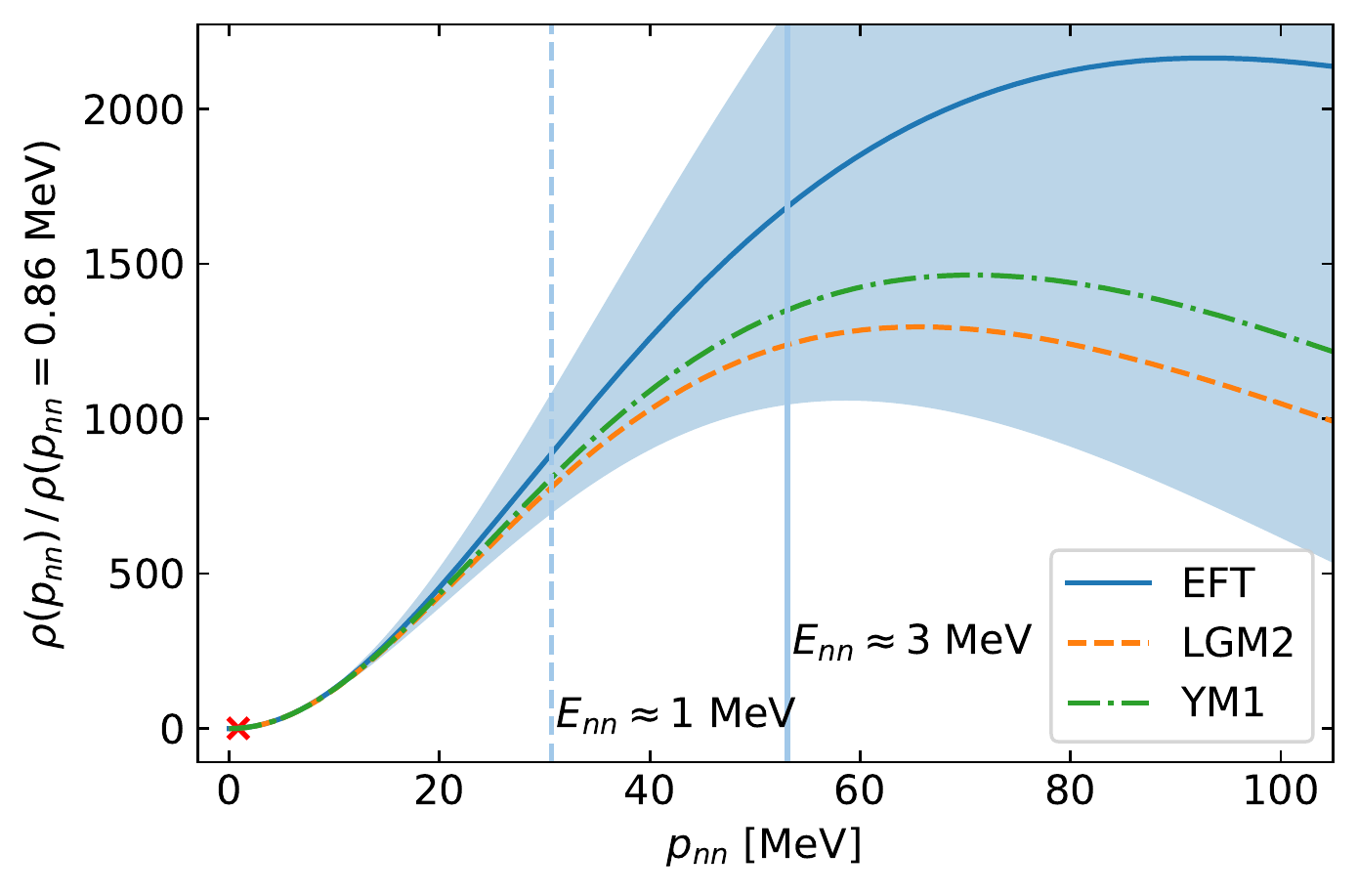}
    \caption{\lgmacronym 2 in comparison with a Yamaguchi model calculation with a three-body force (YM1) and with an EFT calculation.
    In all three cases, the three-body force is tuned
    to the physical binding energy.}\label{fig:tyg_face}
\end{figure}

In preparation for comparing different variants of YM with Halo EFT, we have to check that the YM calculations
yield results similar to our \lgmname s.
\Cref{fig:tyg_face} shows \lgmacronym 2 in comparison with YM1 and an EFT calculation, with the three-body forces of all three calculations tuned
to the physical binding-energy.
\lgmacronym 2 was chosen because it has the same interaction channels as YM1 and the LO EFT for \hesix.

It can be seen that the discrepancy between \lgmacronym 2 and YM1 is much smaller than the one between \lgmacronym 2 and the EFT.
This implies, that if one understands roughly the discrepancy between YM1 and EFT, one probably also understands much of the
difference between \lgmacronym 2 and EFT.
Therefore, we proceed with comparisons of different variants of the Yamaguchi model with the EFT.

\begin{figure}[H]
    \centering
    \includegraphics[width=.42\textwidth]{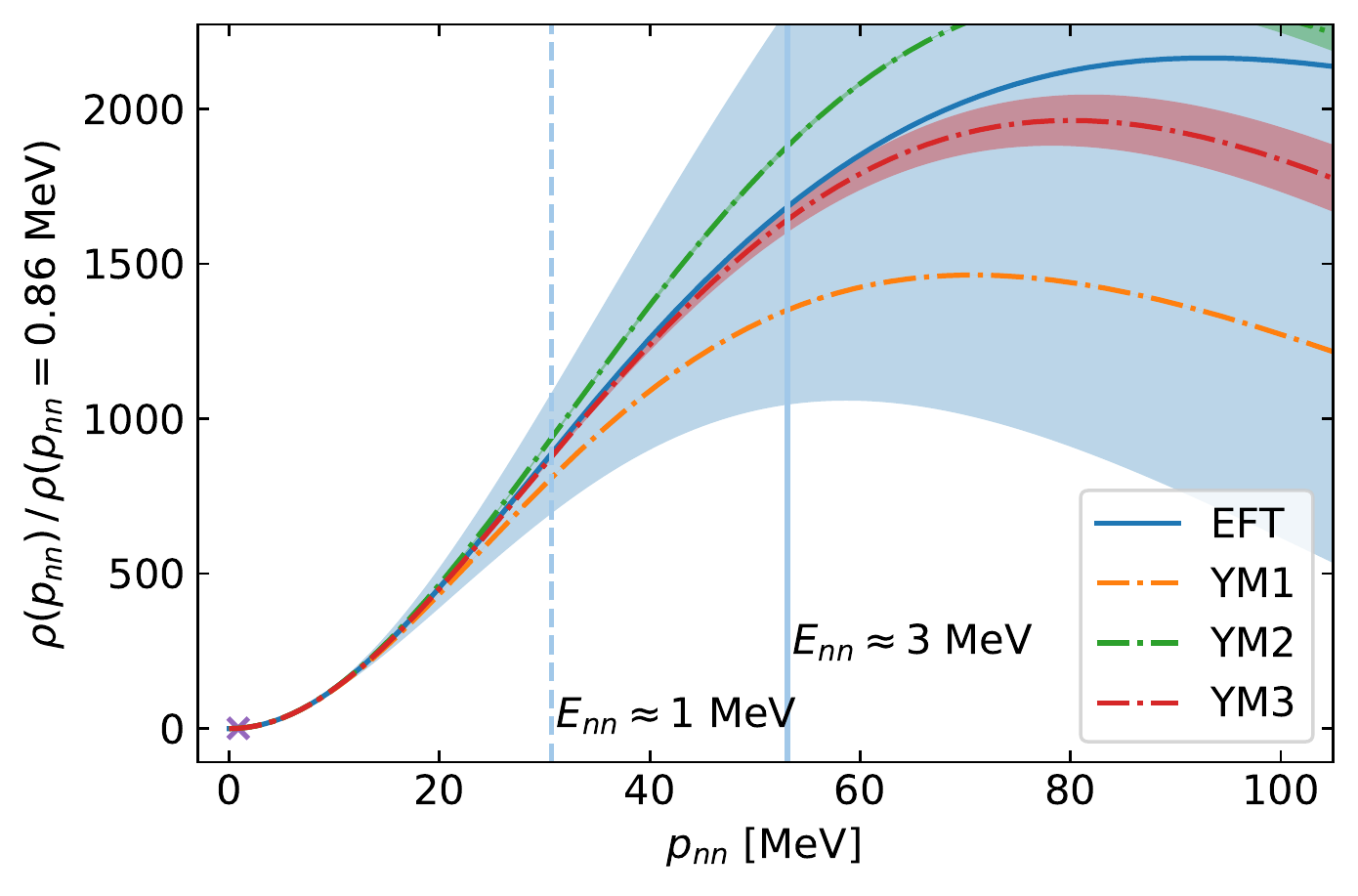}
    \caption{YM results in comparison with the Halo EFT result.
    All are normalized to have a certain arbitrary value at a
    momentum indicated by the red cross. Meanwhile, 
    the vertical lines indicate relative energies of \SI{1}{\mev} and \SI{3}{\mev}.
    The estimated numerical uncertainties of the YM results are indicated with uncertainty bands.
    The estimation is based on the comparison of the calculation with a three-body cutoff at momenta of
    \SI{2250}{\mev} with one with a cutoff of \SI{1500}{\mev} and half as many mesh points. The error band for the 
    Halo EFT result is generated using the expected size of the NLO correction. 
    }\label{fig:tyg_eft}
\end{figure}

\Cref{fig:tyg_eft} shows the three YM calculations in comparison with the EFT calculation.
It can be seen  that YM2 agrees better with the EFT curve than YM1.
This shows that the unitary term of the \(nc\) interaction, which is not present in the LO Halo EFT calculation, has a remarkable influence.
The agreement of YM3, where additional higher order terms are also removed, with EFT is even better.
As a conclusion, we can say that higher order terms of the \(nc\) interaction have a significant influence on
the distribution at higher momenta.
While we can expect improved agreement of an NLO Halo EFT calculation with \lgmacronym~because of the EFT principles, we have now
seen specific effects, in particular the unitarity piece of the \(nc\) t-matrix, that will make an NLO result agree better with model calculations. 

\section{\texorpdfstring{Influence of the \(\boldsymbol{nn}\) scattering length on the ground-state \(\boldsymbol{nn}\) relative-energy distribution}{Influence of the nn scattering length on the ground-state nn relative-energy distribution}}
\label{sec:gs_distrib_nn_sl}
\begin{figure}[H]
  \centering
  \includegraphics[width=.5\textwidth]{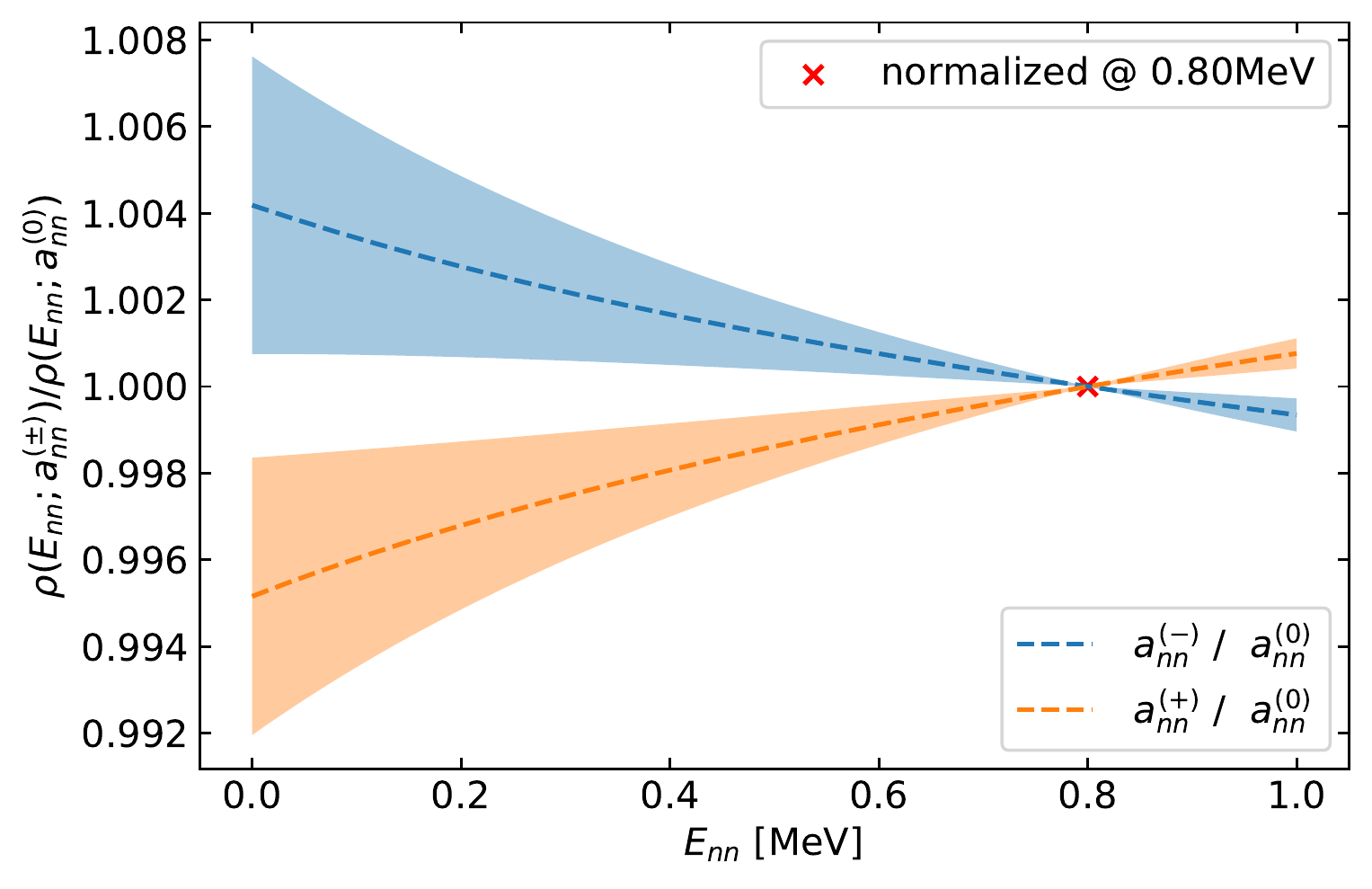}
  \caption{Ratios of ground-state \(nn\) relative-energy distributions obtained with different \(nn\) scattering lengths.
  The definitions \(a_{nn}^{(+)} = \SI{-16.7}{\fm}\), 
  \(a_{nn}^{(0)} = \SI{-18.7}{\fm}\) and \(a_{nn}^{(-)} = \SI{-20.7}{\fm}\) hold.
  All results are based on \(\Psi_c{\K{p,q}}\).
  \lbh~was used. Uncertainty bands based on comparison with calculation with
  half as many mesh points and \lbl~are plotted.
  As the variations in the ratio with scattering length are less than 1\%, the small uncertainties associated with mesh-point 
  and cutoff choices 
  result in quite large bands in this plot.  
  The underlying distributions are normalized to a value of 1 at \(E_{nn}\approx \SI{0.8}{\mev}\), therefore
  the quotients are 1 at this point.
  }\label{fig:gs_en_distribs_rel_comp}
\end{figure}

\Cref{fig:gs_en_distribs_rel_comp} shows the influence of the \(nn\) scattering length on the ground-state (i.e., no 
final-state interaction) relative-energy distribution.

It can be seen that the influence of the scattering length on the ground-state distribution is quite small.
The discussion of \hcref{Appendix B of the main part} regarding the uncertainty bands applies analogously.

\section{Summary of two-potential scattering theory results we use}\label{sec:two_pot_scat_theory}
The purpose of this section is to review the findings of two-potential scattering theory used in
\hcref{Section IV of the main part of the paper} in a bit more detail.
The discussion presented here is based on that in Ref. \cite{goldberger2004collision}.
The two potentials are given by the production potential \(V\) and 
the potential \(U\) causing the final-state interactions.
We make use of a helpful identity of two-potential scattering theory
for calculating the probability amplitude of the transition from a state \(\ket{\alpha}\) to a state \(\ket{\beta}\):
\begin{equation}
  T_{\beta \alpha} = \mel{\beta}{T_{U+V}^{(+)}}{\alpha} \,.
\end{equation}
These states satisfy the equations \(H_0 \ket{\alpha} = E_\alpha \ket{\alpha}\) and \(H_0 \ket{\beta} = E_\beta \ket{\beta}\) 
with \(E_\alpha = E_\beta = E\). 
\(H_0\) should be thought of as the part of the Hamiltonian that does not
include the interactions \(U\) and \(V\).
The operator \(T_{U+V}^{(+)}\) is then the t-matrix for scattering involving \(U\) and \(V\). 
It satisfies the Lippmann-Schwinger equation
\begin{equation}
  T_{U+V}^{(+)} = \K{U + V} + \K{U+V} G_0^{(+)} T_{U+V}^{(+)} \,.
\end{equation}
We follow the approach of Ref. \cite{goldberger2004collision} and dissect this overall t-matrix using the M{\o}ller operators that correspond to the potentials \footnote{
  The general M{\o}ller operator with a general \(H_0\) having orthonormal eigenstates \(\ket{\phi_{\xi}}\)
  with energies \(E_{\xi}\) reads
  \mbox{\( \Omega_V^{(\pm)} = \id + \int \dd \xi \K{ E_{\xi} - H_0 - V \pm \ci \epsilon  }^{-1} V \ketbra{\phi_{\xi}}{\phi_{\xi}}\).}
  E.g., in the case of a two-body system with \mbox{\(H_0 = \v{p}^2 / \K{2\mu}\)} the M{\o}ller operator reads
  \mbox{\( \Omega_V^{(\pm)} = \id + \int \dd^3{p} \K{ E_{\v{p}} - H_0 - V \pm \ci \epsilon  }^{-1} V \ketbra{\v{p}}{\v{p}}\).}
  The latter formula can be found in Ref. \cite{taylor2006scattering}.
  Since we assume in our discussion that the M{\o}ller operator acts on the eigenstate with energy \(E\),
  no integral and no projection operator is needed.
}:
\begin{equation}\label{eq:def_moller_op}
  \Omega_V^{(\pm)} = \id + \K{ E - H_0 - V \pm \ci \epsilon  }^{-1} V \,.
\end{equation}
Employing these operators we can rewrite $T_{U+V}^{(+)}$ as
\begin{equation}
  T_{U+V}^{(+)} = \K{\Omega_U^{(-)}}^\dagger V \Omega_{U+V}^{(+)} + \K{\Omega_U^{(-)}}^\dagger U \,.
\end{equation}
The second term only contributes for elastic scattering reactions as \(V\), 
which causes the production of the final state, is missing there.
We now specialize to cases in which \(V\) induces a transition to an eigenstate of $H_0$ that is orthogonal to the initial eigenstate \(|\alpha \rangle\), e.g.,
because this new overall state contains a subsystem that has a different energy than the initial energy of this subsystem.
This is the case if \(V\) induces the transition from a bound state to a scattering state in a subsystem.
Therefore we have
\begin{equation}\label{eq:Tba_only_first_term}
  T_{\beta \alpha} = \mel{\beta}{ \K{\Omega_U^{(-)}}^\dagger V \Omega_{U+V}^{(+)} }{\alpha} \,.
\end{equation}

We now modify the previous discussion to accommodate the case that 
the final-state interaction \(U\) is
part of the Hamiltonian that describes the initial state.
In such a case the stationary Schr\"odinger equation for the initial state becomes \(\K{H_0 + U} \ket{\alpha} = E_\alpha \ket{\alpha}\).
We continue to assume a free final state, so \(H_0 \ket{\beta} = E_\beta \ket{\beta}\) stays unchanged.
Goldberger and Watson show in Ref. \cite{goldberger2004collision}, that in this case the expression (\ref{eq:Tba_only_first_term}) can be recast as
\begin{equation}\label{eq:Tba_only_first_term_sc_1}
  T_{\beta \alpha} = \mel{\beta}{ \K{\Omega_U^{(-)}}^\dagger V \K{  \id + \K{E-K-U-V + \ci \epsilon}^{-1} V } }{\alpha} \,,
\end{equation}
i.e., our previous result from \cref{eq:Tba_only_first_term} with the right-most \(U\) of the \(\Omega_{U+V}^{(+)}\) removed.
This result can also be obtained from \cref{eq:Tba_only_first_term} with a
redefinition of \(U\) given by \(U \rightarrow  \K{\id - \ketbra{\alpha}{\alpha}} U \K{\id - \ketbra{\alpha}{\alpha}} \).

With \cref{eq:Tba_only_first_term_sc_1} we have arrived at the relation which is the starting point for the direct
calculation of \(nn\) FSI as well as the derivation of the enhancement factors.

\section{Details on the FSI enhancement factors}\label{sec:FSI_EFs}
  In this section we briefly sketch the derivation of FSI enhancement factors and also discuss the multiple options available for enhancement factors.
  The starting point for the derivation is \cref{eq:Tba_only_first_term_sc_1}.
  We consider a final state of definite momentum: \(\ket{\beta} = \ket{\v{p}}\).
    The initial state is the bound state \(\ket{\Psi}\).
    The expression for the transition amplitude \(\mathcal{T}\)
    then reads
    \begin{equation}
      \mathcal{T} = \mel{ \Phi_{\v{p}}^{(-)} }{ V\K{\id + \K{E-K-U-V + \ci \epsilon }^{-1} V} }{\Psi} \,,
    \end{equation}
    where \(\bra{\Phi_{\v{p}}^{(-)}}\) is the scattering state corresponding to a plane wave of momentum \(\v{p}\)
    as incoming wave.
    If the production potential \(V\) is weak, taking only the lowest order of the modified M{\o}ller operator\footnote{
      We call it ``modified" as the full M{\o}ller operator itself would have \(U+V\) instead of \(V\) on the right, see
      \cref{eq:def_moller_op} with \(V \to U + V\).
    }
    on the right is already a good approximation: 
    \(\mathcal{T} \approx \mel{ \Phi_{\v{p}}^{(-)} }{ V }{\Psi}\).
    Evaluating this expression for a local production potential in coordinate space and taking only
    \(s\)-waves into account yields
    \begin{equation}\label{eq:fsi_pa_r_int}
      \mathcal{T} \propto \int \dd r \, r^2 \K{ \Phi_p^{(-)}{\K{r}} }^* V(r) \Psi(r) \,.
    \end{equation}
    For short distances the scattering wave function \(\Phi_p^{(-)}{\K{r}}\) can be approximated
    using the Jost function \(D(E)\): 
    \begin{equation}
      \Phi_p^{(-)}{\K{r}} \approx \mathcal{N} \K{(D(E))^*}^{-1} \sin\K{pr}/\K{pr} \,,
    \end{equation}
    where \(E=p^2 / (2\mu)\) holds. \(\mathcal{N}\) is a normalization constant.
    This approximation becomes exact at \(r=0\).
    If the wave function of the initial state or the production potential are sufficiently short-ranged, 
    this is a good approximation for the important part of the integral (\ref{eq:fsi_pa_r_int}) and the relation 
    \(\mathcal{T} \propto \K{D(E)}^{-1} \int \dd r \, r^2 V(r) \Psi(r) \sin\K{pr}/\K{pr} \)
    holds.
    
    To gain further insight into this result we deviate slighlty from Ref. \cite{goldberger2004collision} and
    switch to momentum space. By using the definition for the
    zeroth spherical Bessel function \(j_0{\K{x}} \coloneqq \sin{\K{x}}/x\) and the orthogonality relation
    for the spherical Bessel functions we obtain for the momentum-space scattering wave function
    \begin{align}
      \braket{p'}{\Phi_p^{(-)}} &= \int \dd r \, r^2 \braket{p'}{r \vphantom{\Phi_p^{(-)}} } \braket{r}{\Phi_p^{(-)}}
                                 \propto \int \dd r \, r^2 \braket{p'}{r}                          \frac{1}{\K{D(E)}^*} j_0{\K{pr}} \nonumber \\
                                &\propto \frac{1}{\K{D(E)}^*} \int \dd r \, r^2 j_0^*{\K{p' r}} j_0{\K{pr}}
                                 \propto \frac{1}{\K{D(E)}^*} \frac{ \de{p'-p} }{p^2} \,.
    \end{align}
    This momentum-space approach makes evaluation of the expression for the transition amplitude straightforward:
    \begin{equation}
      \mathcal{T} \approx \mel{\Phi_p^{(-)}}{V}{\Psi} = \braket{\Phi_p^{(-)}}{\tilde{\Psi}} 
      = \int \dd p' p'^2 \braket{\Phi_p^{(-)}}{p'} \braket{p'}{\tilde{\Psi}}
      \propto \frac{1}{D(E)} \tilde{\Psi}{\K{p}} \,,
    \end{equation}
    where we replaced \(V\ket{\Psi}\) as \(\ket{\tilde{\Psi}}\) for simplicity without loss of generality.
    Squaring, we find for the transition probability the expression
    \begin{equation}\label{eq:fsi_pa_factorization_G1}
      |\mathcal{T}|^2 \propto |D(p^2/\K{2\mu})|^{-2} | \tilde{\Psi}{\K{p}} |^2 \,.
    \end{equation}
    The first factor is usually called the FSI enhancement factor.
    In this case it is the factor from Ref. \cite{goldberger2004collision}:
    \begin{equation}
      G_1{\K{p}} \coloneqq |D(p^2/\K{2\mu})|^{-2} \,.
      \label{eq:G1result}
    \end{equation}

    This derivation shows that it is possible to obtain different enhancement factors by inserting
    different approximations for the scattering wave function in Eq.~(\ref{eq:fsi_pa_r_int}).
    If the production potential or the initial wave function entering there peaks at some \(\tilde{r}\), 
    we search for an approximation of the scattering wave function \(\Phi_p^{(-)}{\K{r}}\)
    that is good around this \(r=\tilde{r}\). If \(\tilde{r}\) is large enough that the effects of the final-state interaction potential 
    \(U\) are small for \(r > \tilde{r}\) then we can make the approximation
    \begin{equation}\label{eq:phi_p_for_Ggen}
      \Phi_p^{(-)}{\K{r}} \approx g{\K{pr}} |_{r=\tilde{r}} \, \frac{ \sin{\K{pr}} }{pr},
    \end{equation}
    with the function \(g\) determined by our approximation,
    the integral over $r$ again picks out a particular momentum in the wave function \(|\tilde{\Psi} \rangle\), and one obtains for the transition probability the relation
    \begin{equation}\label{eq:fsi_pa_factorization_Ggen}
      |\mathcal{T}|^2 \propto G_{g,\tilde{r}}{\K{p}} | \tilde{\Psi}{\K{p}} |^2 \,,
    \end{equation}
    with the enhancement factor now given by
    \begin{equation}\label{eq:g_and_Ggen}
      G_{g,\tilde{r}}{\K{p}} = | g{\K{p\tilde{r}}} |^2 \,.
    \end{equation}
    This result simplifies the derivation of additional enhancement factors and determines also how
    these have to be used.
 
    If we assume that the production potential \(V\) does (almost) not change the momentum \(p\),
    we can replace \(\tilde{\Psi}\) in \cref{eq:fsi_pa_factorization_Ggen} (or \cref{eq:fsi_pa_factorization_G1} respectively)
    by \(\Psi\) and obtain
    \begin{equation}
      \rho^{(G_i)}{\K{p}} \propto G_i{\K{p}} \rho{\K{p}}\,,
    \end{equation}
    where we write \(\rho\) for the probability distribution given by the absolute square of the wave function times
    measure factors.
    Note, that this relation does not determine the absolute value of the distribution but its momentum dependence, as
    we have already neglected normalization factors at some points.    If the knockout reaction would have 100\% probability, then the absolute values of the final probability distribution
      would be fixed by the conservation of probability. However, there are also other final states we do not observe and in order to
      predict the amount of probability of producing the final state of interest more detailed knowledge of the production potential would be necessary.

    Up to this point, the discussion of FSI enhancement factors was specific to two-body systems.
    The underlying two-potential scattering theory holds generally for \(n\)-body systems, as can be seen from our discussion.
    Therefore, the derivation of FSI enhancement factors can be generalized to \(n\)-body systems, if the FSI happens (approximately) only in a certain two-body subsystem (see, e.g., Refs. \cite{Watson:1952ji,goldberger2004collision}).
    The considered final state of this subsystem would then be a state of definite momentum.
    Since the FSI enhancement factor factorizes the FSI from the action of the production potential, it can be used not only
    in the case where the production potential acts in the same subsystem as the FSI potential,
    but also in cases where the two act on different subsystems of the overall \(n\)-body system.
    The requirement for introducing the enhancement factor is that the coordinate-space \(\tilde{\Psi}\) peaks sharply enough in the subsystem such that this approximation can be made.
    
    This derivation and discussion shows that enhancement factors are a viable way to treat \(nn\) FSI in the reaction ${}^6{\rm He}(p,p\alpha)nn$. We thus now turn to a 
     brief review of different \(nn\) FSI enhancement factors that have been used in the literature. 
    The explicit expression for the Jost-function based enhancement factor \(G_1(p)\) derived in Ref. \cite{goldberger2004collision}
    is given by\footnote{
      Note, that in Ref. \cite{goldberger2004collision} the enhancement factor has \(1/a_{nn}\) instead of \(-1/a_{nn}\)
      in the denominator. This is rooted in a different sign convention for the scattering length.
      We define \(k\cot{\K{\delta_0{(k)}}} = -1/a_0 + r_0 k^2  /2 + \mathcal{O}{\K{k^4}}\).
    }
    \begin{equation}\label{eq:G1}
      G_1{\K{p}} = \frac{ \K{\K{p^2 + \alpha^2} r_{nn}/2 }^2 }{ \K{-a_{nn}^{-1} + \frac{r_{nn}}{2}p^2 }^2 + p^2 } \,,
    \end{equation}
    where \(\alpha\) is given by \(1/r_{nn} \K{1 + \sqrt{1 - 2 r_{nn}/a_{nn}} } \).
    One advantage of this expression is the right asymptotic behavior for \(p \to \infty\).
    The relation \(\lim_{p \to \infty} G_1{\K{p}} = 1\) is fulfilled, meaning that for infinitely high energies the
    effect of the final-state interaction becomes negligible.
    
    There is also another enhancement factor given in Ref. \cite{boyd69}, which reads
    \begin{equation}\label{eq:G2}
      G_2{\K{p}} = \frac{ \K{a_{nn}^{-1} - \frac{r_{nn}}{2} p^2 -r_{nn}^{-1} }^2 }{ \K{a_{nn}^{-1} - \frac{r_{nn}}{2} p^2}^2 + p^2 } \,.
    \end{equation}
    It can been seen as a special case of our more general FSI enhancement factor \(G_{\tilde{r}}\),
    which has a continuous parameter \(\tilde{r}\) describing the scale at which the production potential is assumed
    to peak sharply and where the final-state interaction is already almost zero.   This enhancement factor is given by
    \begin{equation}\label{eq:Gtr}
      G_{\tilde{r}}{\K{p}} = \frac{ \K{ a_{nn}^{-1} - \frac{r_{nn}}{2} p^2  - \tilde{r}^{-1} }^2 }{  \K{a_{nn}^{-1} - \frac{r_{nn}}{2} p^2}^2 + p^2  } \,.
    \end{equation}
    Its full derivation can be found in \cref{ap:G_rtilde}.
  \(G_{\tilde{r}}\) and \(G_2\) have the correct asymptotic behavior for \(p \to \infty\).
     Note that it is not sensible to use \(G_{\tilde{r}}\) in the limit \(\tilde{r} \to 0\),
    as it diverges there, and the assumptions invoked in its derivation cease to be valid.
    If one assumes that the production potential peaks at (almost) zero range, we recommend using \(G_1\),
    as it is designed for this case. The derivation uses the exact wave function at \(r=0\), while the derivation 
    of \(G_2\) laid out in \cref{ap:G_rtilde} uses the asymptotic form of the wave function \(\Phi_p^{(-)}(r)\) to set the value of \(g(p\tilde{r})\). This only makes sense if $\tilde{r}$ is greater than the range of the potential.

    \subsection{FSI enhancement factors and the direct calculation}
    Because we also derived the FSI enhancement factors from two-potential scattering theory
    it is straightforward to make the connection between them and the direct calculation of FSI. 
    The FSI enhancement factor \(G_{\tilde{r}}\) from \cref{eq:Gtr}, which can be seen as a generalization of the factor \(G_2\),
    can be obtained from \hcref{Eq. (36) of the main part}.
    To do that the limit \(\Lambda \to \infty\) is taken and the integral approximated as
    \begin{equation}
      \intdd{\pp} \frac{\pp[2] \Psi_c{\K{\pp,q}} - p^2 \Psi_c{\K{p,q}}}{ p^2 - \pp[2] } \approx - \frac{\pi}{2} \tilde{r}^{-1} \Psi_c{\K{p,q}} \,.
    \end{equation}
    In this case the FSI becomes completely independent of the momentum \(q\) and we obtain for the final wave function
    \begin{equation}
      \Psi_c^{\K{G_{\tilde{r}}}}{\K{p,q}}
      = \K{ 1 + \frac{ -\tilde{r}^{-1} - \ci p }{a_{nn}^{-1} - \frac{r_{nn}}{2} p^2 + \ci p} } \Psi_c{\K{p,q}} \,.
    \end{equation}
    The corresponding FSI enhancement factor reads
    \begin{equation}
      \left| 1 + \frac{ -\tilde{r}^{-1} - \ci p }{a_{nn}^{-1} - \frac{r_{nn}}{2} p^2 + \ci p} \right|^2
      = G_{\tilde{r}}{\K{p}} \,.
    \end{equation}
    This provides an alternative interpretation of the FSI enhancement factor: we now see that it corresponds to assuming that the real part of the integral
    in \hcref{Eq. (35) of the main part} can be approximated by \(\Psi_c(p,q)\) times a number of order \(1/\tilde{r}\). 

    \subsection{A derivation for a continuous FSI enhancement factor}\label{ap:G_rtilde}
    On the basis of \cref{eq:phi_p_for_Ggen} and of \cref{eq:g_and_Ggen} we derive a more generic FSI
    enhancement factor for the case that the production potential \(V\) in \cref{eq:fsi_pa_r_int} peaks sharply
    at some \(r=\tilde{r}\).
    Additionally, we assume that this radius \(\tilde{r}\) is outside of the range of the FSI potential.

    We obtain an approximation for the \(s\)-wave part of the scattering wave function, \(\Phi_{p}^{(-)}{\K{r}}\), 
    which we abbreviate as \(\Phi_0{\K{r}}\).
    The general expression for the \(s\)-wave scattering wave function outside the range of the potential reads
    \begin{equation}
      \Phi_0^*{\K{r}} \propto \frac{1}{2\ci k} \Ke{ \K{ \vphantom{I'} 1 + 2 \ci k f_0{\K{k}} } \frac{e^{\ci kr}}{r} - \frac{e^{-\ci kr}}{r} } \,,
    \end{equation}
    where \(f_0{\K{k}}\) is the \(s\)-wave partial wave scattering amplitude given by \(\K{k\cot{\K{\delta_0(k)} - \ci k}}^{-1}\).
    A formal manipulation yields
    \begin{equation}
      \Phi_0^*{\K{r}} \propto \frac{\sin{\K{kr}}}{k r} \K{\vphantom{I'} 1 + \ci k f_0{\K{k}} + k f_0{\K{k}} \cot{\K{kr}} } \,,
    \end{equation}
    which fits the pattern described in \cref{eq:phi_p_for_Ggen}.
    Based on \cref{eq:g_and_Ggen} we obtain the enhancement factor
    \begin{equation}
      \left| 1 + \ci k f_0{\K{k}} + k f_0{\K{k}} \cot{\K{k\tilde{r}}} \right|^2 \,.
    \end{equation}
    Finally, we insert the effective range expansion up to second order for \(f_0(k)\) and make the approximation
    \(k\cot{\K{k\tilde{r}}} \approx 1/\tilde{r} \), which holds for small \(k\tilde{r}\):
    \begin{equation}
      \left| 1 + \ci k f_0{\K{k}} + k f_0{\K{k}} \cot{\K{k\tilde{r}}} \right|^2 
      \approx \frac{ \K{ -1/a_0 + r_0 k^2 /2 + k\cot{\K{k\tilde{r}}} }^2 }{ \K{ -1/a_0 + r_0 k^2 /2 }^2 + k^2  }
      \approx  \frac{ \K{ -1/a_0 + r_0 k^2 /2 + 1/\tilde{r} }^2 }{ \K{ -1/a_0 + r_0 k^2 /2 }^2 + k^2  } \eqqcolon G_{\tilde{r}}{\K{k}}\,.
    \end{equation}

    \section{Auxiliary calculations for the FSI formula}\label{ap:fsi_formula}
    The starting point for this appendix is the following integral:
    \begin{equation}
      I{\K{p, q}} \coloneqq \rint{\pp} \reg{{}}{\pp} \K{ p^2 - \pp[2]  + \ci \epsilon }^{-1}  \Psi_c{\K{\pp,q}} \,,
    \end{equation}
    which appears with \(l=0\) in \hcref{Eq.~(35) of the main part}.
    We can rewrite this integral as
    \begin{align}
      I{\K{p, q}} &= \int \dd \pp \frac{ \reg{{}}{\pp} \pp[2] \Psic{\pp, q} - \reg{{}}{p} p^2 \Psic{p,q} }{p^2 - \pp[2]} + \reg{{}}{p} p^2 \Psic{p,q} \int \dd \pp \frac{1}{p^2 - \pp[2] +\ci \epsilon} \nonumber \\
      &= \int \dd \pp \frac{ \reg{{}}{\pp} \pp[2] \Psic{\pp, q} - \reg{{}}{p} p^2 \Psic{p,q} }{p^2 - \pp[2]} - \ci \frac{\pi}{2} p \reg{{}}{p} \Psic{p,q} 
      \eqqcolon I_1{\K{p,q}} - \ci \frac{\pi}{2} p \reg{{}}{p} \Psic{p,q} \,.
    \end{align}
    However, the first integral \( I_1{\K{p,q}}\), which shall be computed numerically, is not an integral over a finite range,
    as the regulator, \(\reg{{}}{p}\) in the second term regulates in \(p\) but not in the integration variable \(\pp\).
    This would complicate the numerical calculation of it.
    We focus on the case of form factors with Heaviside functions as regulators:
    \begin{equation}
      \reg{{}}{p} = \Theta{\K{\Lambda - p}} p^l \,.
    \end{equation}
    In this case, we can circumvent this difficulty.
    We split this first integral into an integral over a finite range, which can be easily calculated numerically,
    and an integral over an infinite range.
    The latter one does not depend on the wave function, thus it can be computed analytically.
    We obtain
    \begin{align}
      I_1{\K{p,q}} &= \int_{0}^\Lambda \dd \pp \frac{ \pp[2] \pp[l] \Psic{\pp, q} - p^2 p^l \Psic{p,q} }{p^2 - \pp[2]}
         - \reg{{}}{p} p^2 \Psic{p,q} \int_{\Lambda}^\infty \dd \pp \frac{ 1 }{p^2 - \pp[2]}  \nonumber \\
      &= \int_{0}^\Lambda \dd \pp \frac{ \pp[2] \pp[l] \Psic{\pp, q} - p^2 p^l \Psic{p,q} }{p^2 - \pp[2]}
         + \frac{1}{2} \reg{{}}{p} p \Psic{p,q} \ln{\K{ \frac{\Lambda + p}{\Lambda - p} }},
    \end{align}
    where we have assumed that \(p < \Lambda\) holds.
    Based on this result we obtain for \(I{\K{p,q}}\) the relation
    \begin{equation}
      I{\K{p, q}} = \int_0^\Lambda \dd \pp \frac{ \pp[2] \pp[l] \Psic{\pp, q} - p^2 p^l \Psic{p,q} }{p^2 - \pp[2]} 
      + \K{ \frac{1}{2} \ln{\K{ \frac{\Lambda + p}{\Lambda - p} }} - \ci \frac{\pi}{2} } \reg{{}}{p} p\Psic{p,q} \,.
    \end{equation}

\section{Momentum-space wave functions from the FaCE results}\label{ap:face_mom_sp_wf}
We describe the calculation of the momentum-space wave function from the FaCE output following Ref. \cite{Chulkov:1990ac} and Ref. \cite{Zhukov:1993aw}.
With FaCE we obtain value tables for the \(\cf{K,l}{S}{\rho}\) functions.
In a first step towards a wave function depending on \(p\) and \(q\) momenta, we calculate
a momentum-space analog of \(\cf{K,l}{S}{\rho}\), i.e. \(\cf{K,l}{S}{\hp}\):
\begin{align}\label{eq:chi_trafo}
  \cf{K,l}{S}{\hp} = \ci^K \hp^{-2} \int_0^\infty \cf{K,l}{S}{\rho} J_{K+2}{\K{\hp\rho}} \sqrt{\rho} \dd \rho \,,
\end{align}
where the cylindrical Bessel function of order \(i\) is given by \(J_i\).

In a next step, we obtain the wave function.
Note that we use in this context a convention for the Jacobi momenta which is different from the one in our EFT calculations.
The convention employed here is used, e.g., in Ref. \cite{Zhukov:1993aw}:
we mark these "new" Jacobi momenta with a tilde.
In the case of the core as spectator the relations between these two sets are given by
\begin{align}\label{eq:p_cv}
    \v{p_c} &= \v{\tilde{p}_{c}} / \sqrt{2}                    \,, \\
    \v{q_c} &= \v{\tilde{q}_{c}} / \sqrt{\K{A_c + 2}/\K{2A_c}} \,, \label{eq:q_cv}
\end{align}
where \(A_c\) is the core-to-neutron mass ratio.
Note that in case of \(\v{p_c}\) and \(\v{q_c}\) we omit the index in many cases completely,
as the function or the bra/ket usually already has an index.
The total wave function containing all partial wave components reads
\begin{align}
    \Psi_c^{(\mathrm{cpl})}{\K{ \v{\tilde{p}_{c}}, \v{\tilde{q}_{c}} }} &= \sum_{K,l} \cf{K,l}{S=0}{\hp} \hsy{K,L=0,0}{l,l}{\v{\tilde{p}_{c}},\v{\tilde{q}_{c}} } \ket{S=0, M_S=0} \nonumber \\
    &\quad+ \sum_{K,l} \cf{K,l}{S=1}{\hp} \Ke{ \hsy{K,L=1,M_L}{l,l}{\v{\tilde{p}_{c}},\v{\tilde{q}_{c}}} \ket{S=1,M_S} }_{J=0,M_J=0}\,, 
\end{align}
where the hyperspherical harmonic is denoted by \(\hsy{K,L,M_L}{l,\lambda}{\v{\tilde{p}_{c}},\v{\tilde{q}_{c}}}\) with the subsystem orbital angular
momentum quantum numbers \(l\) and \(\lambda\) and the total orbital angular momentum quantum number \(L\) and its projection \(M_L\).
\(\alpha\) is the hyperangle given by \(\alpha \coloneqq \tan{\K{\v{\tilde{p}_{c}} / \v{\tilde{q}_{c}}}} \).
The rectangular bracket denotes the coupling of the states to the specified \(J\) and \(M_J\).
The momentum \(\hp\) is the hypermomentum\footnote{
  Nomenclature for \(K\) and the momenta may deviate slightly from the one by Zhukov \textit{et al.}.
  In Ref. \cite{Zhukov:1993aw} the momentum \(\v{\tilde{p}_c}\) is denoted by \(\v{p_3}\) and \(\v{p_{nn}}\).
  The momentum \(\v{\tilde{q}_c}\) is denoted by \(\v{q_3}\) and \(\v{p_c}\).
}
defined as \(\hp^2 \coloneqq \tilde{p}_{c}^2 + \tilde{q}_{c}^2\).

The hyperspherical harmonics are given by
\begin{align}
    \hsy{K,L,M_L}{l,\lambda}{\v{\tilde{p}_{c}},\v{\tilde{q}_{c}}} \coloneqq \Phi_K^{l,\lambda}{\K{\alpha}} \cy{l,\lambda}{L,M_L}{\v{\tilde{p}_{c}},\v{\tilde{q}_{c}}}\,,
\end{align}
where the coupled spherical harmonic is denoted by \(\cy{l,\lambda}{L,M_L}{\v{\tilde{p}_{c}},\v{\tilde{q}_{c}}}\).
The function \(\Phi\) is defined by
\begin{align}
    \Phi_K^{l,\lambda}{\K{\alpha}} &\coloneqq N_K^{l,\lambda} \K{\sin{\alpha}}^l \K{\cos{\alpha}}^\lambda P_n^{l+1/2,\lambda+1/2}{\K{\cos{\K{2\alpha}}}}\,, \\
    N_K^{l,l} &\coloneqq \frac{ \sqrt{ \K{2K+4} \K{K/2-l}! \K{K/2+l+1}! } }{ \Gamma{\K{\K{K+3}/2}} } \,,
\end{align}
where \(n=\K{K-l-\lambda}/2\) holds.
The Jacobi polynomial is denoted by \(P_n^{a,b}\).

The wave-function component relevant for our investigations is given by
\begin{equation}
  \Psi_c{\K{\tilde{p},\tilde{q}}} = \ibraket{c}{ \tilde{p},\tilde{q}; \Omega_c^{(0,0,0)} }{ \Psi }{}
  = \sum_K \cf{K,l=0}{S=0}{\hp} \Phi_K^{0,0}{\K{\alpha}} \frac{1}{4\pi} \,.
\end{equation}

\section{\texorpdfstring{Removal of the \(\boldsymbol{\poh}\) \(\boldsymbol{n\alpha}\) interaction in FaCE}{Removal of the 2P1/2 n alpha interaction in FaCE}}\label{ap:turn_poh_off}
We describe here how the central and the spin-orbit potential depths are tuned in order to turn the \(n\alpha\) interaction in the \(\poh\) partial wave off
while leaving the one in the \(\pth\) partial wave unchanged.
We make use of the fact that the central and spin-orbit potentials have the same range parameter.
Using \(\v{L} \cdot \v{S} = \frac{1}{2}\K{ \K{\v{L}+\v{S}}^2  - \v{L}^2 - \v{S}^2 }\) we obtain for the combination of central and spin-orbit
potential in \poh~and \pth
\begin{align}
  V^{(\pth)}{\K{r}} &= \K{ \frac{1}{2} \bar{V}_{SO}^{(1)}  + \bar{V}_{c}^{(1)} }\exp\K{-r^2/\K{a_{1}^2}}\,,\\
  V^{(\poh)}{\K{r}} &= \K{ - \bar{V}_{SO}^{(1)} + \bar{V}_{c}^{(1)} }\exp\K{-r^2/\K{a_{1}^2}}\,,
\end{align}
where we omitted the index specifying the potential type in the range parameter, since this parameter is identical for both
types of potentials.
These equations make it clear that in order to turn the interaction in \poh~off, one need only set \(\bar{V}_{SO}^{(1)} = \bar{V}_{c}^{(1)}\).
By adopting \(\bar{V}_{c}^{(1)}=\SI{-35.45}{\mev}\), one then obtains an unchanged overall depth in the \pth~channel and zero potential in the \poh\ channel.

\section{FaCE settings used for \lgmacronym~calculations}\label{ap:face_num_param}
When checking the numerical and the model space convergence of the FaCE results, several
aspects must be taken into account.
First, there is a parameter \(K_\mathrm{max}\) for the FaCE program. It determines the maximum \(K\) for which
\(\chi_{K,l}^S\) is calculated.
FaCE has also additional numerical parameters such as the step width and maximum value of the hyperradius for tabulating
the \(\chi_{K,l}^S{\K{\rho}}\).
The calculation of the momentum distribution from the FaCE results also has model space and numerical integration parameters
which have to be checked in the convergence analysis.
One parameter is \(K_\mathrm{trunc}\), that is \(\leq K_\mathrm{max}\) and determines which wave-function components are used in the calculation of the momentum distribution. 
The convergence of the FaCE in all these parameters 
is checked for each calculation. 

\Cref{tab:face_calc_parameters} lists the used values for the most important computational parameters of FaCE
and the subsequent calculations.
Two different parameter sets corresponding to two different accuracy levels were used in order to estimate the computational uncertainties of the results.
The relative deviation between the results obtained with lower and the higher accuracy level (sp vs. hp) 
for \lgmacronym 1, \lgmacronym 2, \lgmacronym 3, \lgmacronym 1SR, and \lgmacronym 2SR is smaller than 2\%. (Note that not all parameters which were varied are listed in the table.)

\begin{table}[H]
  \centering
  \caption{This table lists the different values for the parameters which were used to determine
  whether the results are numerically converged. These parameters are the ones of FaCE or subsequent calculations for obtaining
  the ground-state \(nn\) relative-momentum distribution. The one parameter set is labeled as "sp" (standard precision),
  the other one is labeled as "hp" (high precision).
  Integration and interpolation boundaries in momentum space are generically denoted with \(p_\mathrm{min}\) and \(p_\mathrm{max}\).
  Note that this table is not complete, but the most important computational  parameters are contained in it.}\label{tab:face_calc_parameters}
  \begin{tabular}{m{3.5cm} ccc}
    \toprule
    & & \textbf{sp} & \textbf{hp} \\
    \hline
    \multirow{6}{*}{\textbf{FaCE}} & \(K_\mathrm{max}\) & 18 & 27 \\
         & \(l_\mathrm{max}\) & 12  & 18   \\
         & \(N_\mathrm{Jac}\) & 40  & 60   \\
         & \(N_\mathrm{Lag}\) & 30  & 60   \\
         & \(rr\)             & 0.3 & 0.225 \\
         & \(nbmax\)          & 30  & 60   \\ \midrule
    \multirow{2}{=}{\textbf{FaCE output: \mbox{table of \(\chi_{K,l}^{(S)}{\K{\rho}}\)}}} & \(\rho_\mathrm{max}\) [fm] & 34.0 & 51.425 \\
                                 & \(\Delta \rho\) [fm]       & 0.25  & 0.187  \\ \midrule
    \multirow{2}{=}{\textbf{obtaining the wave function}} & \(K_\mathrm{trunc}\) & 16 & 24 \\ 
                                                          & & & \\ \midrule 
    \multirow{2}{=}{\textbf{interpolation of \(\chi_{K,l}^{(S)}{\K{p}}\) }} & \(p_\mathrm{min}\) [MeV] & 0.001 & 0.0001 \\ 
                                                                            & \(p_\mathrm{max}\) [MeV] & 710   & 1057.5   \\ 
                                                                            & \(N_\mathrm{mesh~points}\) & 71 & 141 \\ \midrule
    \multirow{3}{=}{\textbf{obtaining the distribution: \(q\)-integration}} & \(p_\mathrm{min}\) [MeV]   & 0.001 & 0.0001 \\
                                                                            & \(p_\mathrm{max}\) [MeV]   & 478   & 829    \\
                                                                            & \(N_\mathrm{mesh~points}\) & 192   & 443    \\
    \bottomrule
  \end{tabular}
\end{table}

\bibliographystyle{apsrev4-2}
\bibliography{supplement_he6_and_nn_scattering_length.bbl}